\documentclass[reprint,showpacs,prd,preprintnumbers,amssymb,nofootinbib,superscriptaddress,aps]{revtex4-1}
\pdfoutput=1
\usepackage{amsmath,amssymb,bm,natbib,mathtools,slashed}
\makeatletter
\let\atop\@@atop
\makeatother
\usepackage{bbold}
\usepackage{epsfig}
\usepackage{hyperref}
\usepackage{graphicx}
\usepackage{subcaption}
\usepackage{csquotes}
\usepackage{color,colordvi,xcolor,soul}
\usepackage{booktabs,multirow,array}
\usepackage{comment,multirow,tabularx,ragged2e,booktabs,caption}
\usepackage{float}
\usepackage{cancel}
\makeatletter
\DeclareRobustCommand\cancel[1]{\ifmmode
	\mathpalette{\@cancel{\@can@slash{}}}{#1}\else 
	\leavevmode\@cancel{\@can@slash{}}\hbox{#1}\fi}
\makeatother

\usepackage[normalem]{ulem}
\renewcommand{\(}{\left(}
\renewcommand{\)}{\right)}
\renewcommand{\{}{\left\lbrace}
\renewcommand{\}}{\right\rbrace}

\newcommand{\del}{\partial}
\newcommand{\nn}{\nonumber}

\newcommand{\order}[1]{\mathcal{O}\({#1}\)}
\newcommand{\msbar}{\overline{\text{MS}}}
\newcommand{\as}{\alpha_\mathrm{s}}

\newcommand{\ordas}[1]{\mathcal{O}\left(\alpha_s^{#1}\right)}
\newcommand{\GeV}{\,\mathrm{GeV}}

\newcommand{\MeV}{\,\mathrm{MeV}}

\newcommand{\hs}{\hspace{.4mm}}

\newcommand{\bs}{\hspace{1cm}}

\begin{document}
	\title{Renormalization group improved determination of light quark masses from Borel-Laplace sum rules.}
	\author{M. S. A. Alam Khan}
	\email{alam.khan1909@gmail.com}
	\affiliation{Centre for High Energy Physics, Indian Institute of Science, Bangalore 560 012, India}
	\begin{abstract}
		We determine masses of light quarks ($m_u$,$m_d$,$m_s$) using Borel-Laplace sum rules and renormalization group summed perturbation theory (RGSPT) from the divergence of the axial vector current. The RGSPT significantly reduces the scale dependence of the finite order perturbative series for the renormalization group (RG) invariant quantities such as spectral function, the second derivative of the polarization function of the pseudoscalar current correlator, and its Borel transformation. In addition, the convergence of the spectral function is significantly improved by summing all running logarithms and kinematical $\pi^2$-terms. Using RGSPT, we find $m_s(2\GeV)=104.34_{-4.21}^{+4.32}\hs\MeV$, and $m_d(2\GeV)=4.21_{-0.45}^{+0.48}\hs\MeV$ leading to $m_u(2\GeV)=2.00_{-0.38}^{+0.33}\hs\MeV$.
	\end{abstract}
	\maketitle
	\section{Introduction}
	The light quark masses are important parameters for quantum chromodynamics (QCD) and electroweak physics. Due to confinement, they are not freely observed, and their values depend on the scheme used. They are taken as input in various quantities related to flavor physics and play a key role in the proton-neutron mass difference and the strong CP violating observable $\epsilon'/\epsilon$, etc. Precise determination of their values has been of constant interest in the past three decades. These masses can be precisely obtained using the lattice QCD simulations, and for recent development, we refer to Ref.~\cite{FlavourLatticeAveragingGroupFLAG:2021npn}.\par 
 Theoretical tools such as the QCD sum rules~\cite{Shifman:1978bx,Shifman:1978by} have played a key role in their precise determination. These sum rules use both theoretical and experimental input on the spectral function and are based on the assumption of the quark-hadron duality~\cite{Poggio:1975af}. On the hadronic side, the spectral functions for the pseudoscalar channel in the case of the strange and non-strange channels do not have experimental data, and therefore, inputs from chiral perturbation theory (ChPT)~\cite{Weinberg:1978kz,Gasser:1983yg,Gasser:1984gg} become very important. For reviews, we refer to \cite{Scherer:2002tk,Scherer:2012zz,Ananthanarayan:2023gzw} and references therein. \par On the theoretical side, operator product expansion (OPE)~\cite{Wilson:1969zs} is used, which has perturbative and non-perturbative contributions. The perturbative corrections are calculated by evaluating the Feynman diagrams, and non-perturbative corrections are the condensates of higher-dimensional operators of quarks and gluons fields. The condensates can be determined from lattice QCD, ChPT, or using QCD sum rules~\cite{Ioffe:2005ym}.\par
Fixed order perturbation theory (FOPT) is the most commonly used prescription in the literature. In this prescription, perturbative series is a polynomial in the strong coupling constant ($\as(\mu)$), quark masses ($m_{q}(\mu)$), and the running RG logarithms ($\log(\mu^{2}/Q^{2})$). The RG invariance of an observable ($\mathcal{O}$), known to a finite order in perturbation theory, is enforced using the RG equation (RGE):
\begin{align}
     \mu^2\frac{d}{d\hs\mu^2}\mathcal{O}=0\,,
     \label{eq:RGE1}
\end{align} 
that results in a cancellation among the coefficients of different orders in $\as$. Solution to Eq.~\eqref{eq:RGE1} can be used to generate the RG logarithms.\par
RGSPT is a perturbative prescription in which the running RG logarithms arising from a given order of the perturbation theory are summed in a closed form to all orders using RGE. As a result, we get an analytical expression for the perturbative series in which  $\as(\mu)\log(\mu^{2}/Q^{2})\sim\order{1}$. This scheme is useful in reducing the theoretical uncertainties arising from renormalization scale dependence. The procedure is described in section~\eqref{sec:rgspt} and some of the applications can be found in Refs.~\cite{Abbas:2012py,Ananthanarayan:2016kll,Ananthanarayan:2022ufx,Ahmady:2002fd,Ahmady:2002pa,Ananthanarayan:2020umo,Ahmed:2015sna,Abbas:2022wnz,Chishtie:2018ipg,Abbas:2022wnz,AlamKhan:2023kgs,AlamKhan:2023dms}.  \par
 The Borel-Laplace sum rule is one of the important method widely used in the literature, especially for the determinations of quark mass~\cite{Jamin:1994vr,Dominguez:1997eu,Jamin:2001zr,Jamin:2006tj,Chetyrkin:2005kn,Dominguez:2007my,Narison:2014vka,Yuan:2017foa,Yin:2021cbb}, and in the extraction of hadronic parameters~\cite{Gelhausen:2013wia,Narison:2014ska,Wang:2015mxa,Narison:2015nxh}, etc. However, the dependence of an unphysical Borel parameter ($u$) and free continuum threshold $s_0$ parameter is present in these determinations. In principle, any determination using this sum rule should be independent of the choice of these parameters, but they are tuned to get reliable results. In addition, the determination of the light quark masses from these sum rules is found to be very sensitive to the renormalization scale, and a linear behavior has been reported in Refs.~\cite{Chetyrkin:2005kn,Dominguez:2014vca}. Also, suppression to hadronic spectral function using pinched kernels~\cite{Dominguez:2007my,Dominguez:2008tt}, mainly used in the finite energy sum rules (FESR), can not be implemented in this sum rules. \par 
With the limitations in hand, our interest in this sum rule is due to two reasons:
 \begin{enumerate}
     \item The formalism developed in the Ref.~\cite{AlamKhan:2023dms} can be used to improve the convergence and reduced renormalization scale dependence for the spectral function by summing kinematical $\pi^2$-terms using RGSPT.
     \item All order summation of the Euler's constant ($\gamma_E$) and Zeta functions arising as a result of the Borel transformation of the RG invariant second derivative of the polarization function using RGSPT.
 \end{enumerate}
It should be noted that these improvements are very crucial and can be used in any Borel-Laplace sum rule-based studies. On the theoretical side, the leading perturbative $\ordas{4}$ corrections to the pseudoscalar two-point function are now available in Ref.~\cite{Gorishnii:1990zu,Chetyrkin:1996sr,Baikov:2005rw} and other OPE corrections from Refs.~\cite{Generalis:1990id,Chetyrkin:1985kn,Generalis:1990id,Jamin:1992se,Jamin:1994vr,Chetyrkin:2005kn}. For low energy region, there is no experimental information for the pseudoscalar spectral density in the resonance region, but it can be modeled using the experimental values of the resonances~\cite{Bijnens:1994ci,Dominguez:1986aa,Maltman:2001gc}. We have used the results of previous studies on hadronic spectral function from Ref.~\cite{Bijnens:1994ci,Dominguez:1986aa,Maltman:2001gc,Schilcher:2013pvu,Dominguez:1997eu} for the strange and non-strange channel.\par
   Hadronic $\tau$ decays are also found to be very useful in the determination of strange quark mass, CKM element $\vert V_{us}\vert$, and strong coupling constant  and more details can be found in Refs.~\cite{Braaten:1991qm,Pivovarov:1991rh,Kambor:2000dj,Maltman:2007ic,Hudspith:2017vew,Boito:2020xli,Ananthanarayan:2022ufx}. These studies use experimental data on the spectral function. Commonly used prescriptions for the perturbative series in these FESR-based studies use FOPT and contour-improved perturbation theory (CIPT). Recently CIPT has been found to be in conflict with the OPE expectations, and for more details, we refer to Refs.~\cite{Hoang:2021nlz,Hoang:2021unk,Benitez-Rathgeb:2021gvw,Benitez-Rathgeb:2022yqb,Benitez-Rathgeb:2022hfj,Gracia:2023qdy,Golterman:2023oml}.  For other light quark mass determinations using sum rules, we refer to Refs.~\cite{Maltman:2001gc,Jamin:2006tj,Dominguez:2014vca,Dominguez:2018azt}.\par 
It should be noted that only the $\msbar$ definition of $\alpha_s$ and $m_{q}$ are used in this article. The value of $\alpha_s(M_z)=0.1179\pm0.0009$ has been taken from PDG~\cite{ParticleDataGroup:2022pth} and evolved to different scales using five-loop $\beta-$function for three flavors using package REvolver~\cite{Hoang:2021fhn} and RunDec~\cite{Chetyrkin:2000yt,Herren:2017osy}. Its value at tau lepton mass ($M_{\tau}$) scale is $\alpha_s(M_{\tau})=0.3139\pm 0.0083$ and has been used in this article. Also, we have used couplant $x(\mu)\equiv\frac{\as(\mu)}{\pi}$ as expansion parameter in the perturbation series and if explicit energy scale is not shown, then $x$ are assumed to be evaluated at renormalization scale $\mu$.\par
In section \eqref{sec:formalism}, we briefly introduce the quantities needed for the Borel-Laplace sum rule. In section~\eqref{sec:rgspt}, we have given a short review of RGSPT. In section~\eqref{sec:hadronic_info}, hadronic parametrization of the spectral function for the strange and non-strange channel is discussed. In section~\eqref{sec:OPE}, OPE contribution and their results in FOPT and RGSPT prescription are discussed. In section~\eqref{sec:ms_md_determination}, results from the previous sections are used for the light quark mass determinations. In section~\eqref{sec:summary}, we give the summary and conclusion of this article, and the supplementary information is provided in appendix~\eqref{app:mass_run} and \eqref{app:dim0adler}.
  \section{Formalism\label{sec:formalism}}
	The current correlator for the divergence of the axial currents is defined as:
		\begin{align}
		\Psi_{5}(q^2)\equiv i \int d^4x\hs e^{i q x} \langle 0|\mathcal{T}\{j_{5}(x) j_{5}^{\dagger}(0)\} |0\rangle\,,
  \label{eq:def_corr}
	\end{align}
	where $j_{5}$ is given by:
	\begin{align}
j_{5}=\partial^{\mu}\left(\overline{q}_1\gamma_{\mu} \gamma_5q_2\right)&=i \left(m_1+ m_2\right)\left(\overline{q}_1\gamma_5 q_2\right)\nonumber\\&=i \left(m_1+ m_2\right)j_{0}\,,
\label{eq:der_J}
	\end{align}
	 and quark masses $m_i$ as well as quark fields $q_i\equiv
	 q_i(x)$ are bare quantities.\par 
  Using Eq.~\eqref{eq:der_J}, the correlation function in Eq.~\eqref{eq:def_corr} after renormalization in the $\msbar$ scheme is related to the pseudoscalar polarization function ($\Pi_P(q^2,\mu^2)$) by relation:
  \begin{align}
      \Psi_{5}(q^2)=\left(m_1+ m_2\right)^2\Pi_P(q^2,\mu^2)\,
      \label{eq:Psi2Pi}
  \end{align}
  where $m_i\equiv m_i(\mu)$. The polarization function $\Pi_P(q^2,\mu^2)$ is given by
  \begin{align}
		\Pi_{P}(q^2,\mu^2)= i \int d^4x\hs e^{i q x} \langle 0|\mathcal{T}\{j_{0}(x) j_{0}^{\dagger}(0)\} |0\rangle\,,
 	\end{align}
where $j_0$ is a renormalized current in the $\msbar$ scheme. Due to the above relation, the sum rule determinations from the correlator in Eq.~\eqref{eq:def_corr} are sometimes known as pseudoscalar determinations. 
  \par
  Using OPE, a theoretical expression for $\Psi_5(q^2)$ is calculated in the deep Euclidean spacelike regions in the limit $m_q^2 \ll q^2$, and the resulting expansion can be arranged as expansion in $1/(q^2)$. At low energies $\sim 1\GeV^2$, instanton effects become relevant, and their contribution is not captured by OPE expansion and therefore are added to it. Further details on the OPE contributions are presented in Sec.~\eqref{sec:OPE}.\par
  	The Borel-Laplace sum rules are based on the double-subtracted dispersion relation for the correlation function. Therefore, it involves the double derivative of $\Psi_5(q^2)$ and the dispersion relation is given by:
	\begin{align}
		\Psi''_{5}(q^2)=\frac{d^2}{d (q^2)^2}\Psi_{5}(q^2)=\frac{2}{\pi}\int_{0}^{\infty}ds\frac{\text{Im}\Psi_{5}(-s-i\hs \epsilon)}{(s-q^2-i\epsilon)^{3}}\,.
		\label{eq:der2Psi}
	\end{align}
The Borel transformation \footnote{We use normalization given in Ref.~\cite{Jamin:1994vr} i.e. $\hat{\mathcal{B}}_{u}[\frac{1}{(x+s)^a}]=\frac{1}{u^a \Gamma[a]}e^{-x/u}$}, with parameter ``u",  is obtained using the Borel operator, $\hat{\mathcal{B}_u}$, defined as:
\begin{align}
	\hat{\mathcal{B}}_u\equiv\lim _{Q^{2}, n \rightarrow \infty \atop Q^{2} / n=u}\frac{(-Q^{2})^n}{\Gamma[n]}\partial_{Q^{2}}^n
 \label{eq:BO}\,,
\end{align}
where, we have used variables $Q^2=-q^2>0$ for the spacelike and $s=q^2>0$ for timelike regions.
 
  Borel parameter $u$ has the dimension of $\GeV^2$ and the Borel transform of Eq.~\eqref{eq:der2Psi} is obtained as:
\begin{align}
	\Psi''_{5}(u)&\equiv\hat{\mathcal{B}}_u\left[\Psi''_{5}(q^2)\right]=\frac{1}{u^3}\hat{\mathcal{B}}_u\left[\Psi_{5}(q^2)\right](u)\nonumber\\&=\frac{1}{\pi u^3}\int_{0}^{\infty} ds\hs e^{-s/u} \hs \text{Im}\Psi_{5}(-s-i\hs\epsilon)\nonumber\\&=\frac{1}{u^3}\int_{0}^{\infty}ds \hs\hs e^{-s/u} \rho_{5}(s)\,,
	\label{eq:bs_rule}
\end{align}
where the spectral density is given by:
\begin{align}
    \rho_{5}(s)=\frac{1}{\pi}\lim_{\epsilon\rightarrow 0}\left[\text{Im}\Psi_{5}(-s-i\hs\epsilon)\right]\,.
    \label{eq:spectral_density}
\end{align}
It should be noted that the value of the $u\gg\Lambda_{QCD}^2$ in $\Psi_{5}''(u)$ is chosen such that higher order terms of the OPE expansion remain suppressed in the Borel transformed OPE expansion.
 \par
The Borel-Laplace sum rules in the RHS of Eq.~\eqref{eq:bs_rule} involve an integration ranging from the low energy regime of the strong interactions to the high energy regime. The spectral density is approximated with the quark hadron-duality. For the low-energy regime, the spectral function is parameterized in terms of pion/kaon poles and resonances present in the channel, and for the high-energy region, results from perturbative QCD are used. The spectral density from these two regimes can be written as:
\begin{align}
    \rho_5(s)=\theta(s_0-s)\rho_5^\text{had}(s)+\theta(s-s_0)\rho_5^{\text{OPE}}(s)\,,
    \label{eq:spectral_decomposition}
\end{align}
where scale $s_0$ separates the two contributions, and its value should be chosen such that the perturbative treatment is justified.\par
Using Eq.~\eqref{eq:spectral_decomposition}, the Borel sum rule in Eq.~\eqref{eq:bs_rule} can be written as:
\begin{align}
		\Psi''_{5}(u)&=\frac{1}{u^3}\int_{0}^{s_0}ds \hs\hs e^{-s/u} \rho^{had}_{5}(s)\nonumber\\&\hs\bs+\frac{1}{u^3}\int_{s_0}^{\infty}ds \hs\hs e^{-s/u} \rho^{OPE}_{5}(s)\,.
		\label{eq:bs_final}
\end{align}
 which is used in this article for the light quark mass determination. \par 
For clarification, various inputs used in Eq.~\eqref{eq:bs_final} are as follows:
\begin{enumerate}
\item \label{item:1}The $\Psi''_{5}(u)$ is obtained from the Borel transformation of $\Psi''(q^2)$, which involves OPE corrections and addition to the instanton contributions. The instanton contributions are small for the choice of the parameters used in this article but relevant as pointed out in Ref.~\cite{Maltman:2001gc}. These contributions are thus obtained using Eq.~\eqref{eq:psi_Borel}+Eq.~\eqref{eq:borel_instal}.
\item \label{item:2} The hadronic spectral density $\rho_{5}^{\text{had}}(s)$ is obtained by the parametrization of the experimental information on the hadrons appearing in the strange and non-strange channels. These constitutions are discussed in section~\eqref{sec:hadronic_info} for non-strange and strange channels, and we use Eq.~\eqref{eq:had_NS} or \eqref{eq:had_S}.
\item \label{item:3}$\rho_5^{\text{OPE}}(s)$ in the RHS of Eq.~\eqref{eq:bs_final} is obtained from the discontinuity of the theoretical expression of the $\Psi_5(q^2)$ which is calculated using the OPE and instanton contributions are also added to it. It has contributions from Eq.~\eqref{eq:rho_exp} +Eq.~\eqref{eq:rho_insta}.
\item Quark mass appears in both sides of Eq.~\eqref{eq:bs_final} except for the integral term containing $\rho^{had}_{5}(s)$.
\end{enumerate}
It should be noted that the main focus of this article is the RG improvement for the theoretical quantities relevant for point~\eqref{item:1} and point~\eqref{item:3} and its impact in the light quark mass determination. 
\section{Review of the RGSPT} \label{sec:rgspt}
In FOPT prescription, a perturbative series $\mathcal{S}(Q^2,\mu^2)$ in pQCD can be written as:
\begin{align}
	\mathcal{S}(Q^2,\mu^2)\equiv \sum_{i=0,j=0}^{j\le i} T_{i,j} x^i L^j \,,
	\label{eq:Pseris}
\end{align}
where $x=\as(\mu)/\pi$ and $L=\log(\mu^2/Q^2)$.
 The RG evolution of the perturbative series in Eq.~\eqref{eq:Pseris} is obtained using its anomalous dimension, $\gamma_S(x)$, by solving:
 
\begin{subequations}
  \begin{align}
	\mu^2 \frac{d}{d\mu^2} \mathcal{S}(Q^2,\mu^2)&=\gamma_S(x) \hs\hs \mathcal{S}(Q^2,\mu^2)\,,
	\label{eq:RGE}\\
		\mu^2 \frac{d}{d\mu^2} x(\mu)&=\beta(x)\,,\label{eq:RGEx}
\end{align}
\end{subequations}
where anomalous dimension $\gamma_S(x)$ and $\beta(x)$ is given by:
\begin{equation}\label{eq:anom_S}
	\begin{aligned}
	\gamma_S(x)&=\sum_{i=0}\gamma_i x^{i+1}\,,\\
	\beta(x)&=\sum_{i=0}\beta_i x^{i+2}\,.	
	\end{aligned}
\end{equation}
Perturbative series in Eq.~\eqref{eq:Pseris} has no large logarithms if we set $\mu^2=Q^2$ and various parameters, such as quark masses and couplings, are evolved to different scales using their RG equations. To account for renormalization scale dependence for a series with vanishing anomalous dimension, we set $\mu^2=\xi \hs Q^2$ and the parameter $\xi $ is often varied in the range $\xi\in \left[1/2,2\right]$. The RG logarithms in Eq.~\eqref{eq:Pseris} still play a key role in canceling the scale dependence arising from other parameters, such as from $\as$ and $m_q$. \par
 In RGSPT, perturbative series in Eq.~\eqref{eq:Pseris} is arranged as follows:
 \begin{align}
 	\mathcal{S}^\Sigma(Q^2,\mu^2)= \sum_{i=0} x^i S_i (x\hs L)\,,
 	\label{eq:summed_ser}
 \end{align}
 where the goal is to obtain a closed-form expression for coefficients:
 \begin{align}
 	S_i (z)=\sum_{j=0}^{\infty} T_{i+j,j} z^j\,,
 	\label{eq:SRcoef}
 \end{align}
where $z\equiv x \hs L\hs$. $S_i (z)$ are functions of one variable where $z\sim\mathcal{O}(1)$. The closed-form solution for them is obtained using RGE. \par
 The RGE in Eq.~\eqref{eq:RGE} results in a set of coupled differential equations for $S_i(z)$, which in compact form can be written as:
\begin{align}
  \left(\sum_{i=0}^{n} \frac{\beta_{i}}{z^{n-i-1}}\frac{d}{d\hs z}\left(z^{n-i} S_{n-i}(z)\right)+\gamma_i S_{n-i}(z)\right)-S'_n(z)=0\,,
\end{align}
The first three coefficients can be obtained by solving the above differential equation and are given by:
 \begin{align}
     S_0(z)=&T_{0,0} w^{-\tilde{\gamma }_0}\,,\nonumber\\ S_1(z)=&T_{1,0} w^{-\tilde{\gamma }_0-1}+T_{0,0} w^{-\tilde{\gamma }_0-1} \Big[(1-w) \tilde{\gamma }_1\nonumber\\ &\bs+\tilde{\beta }_1 \tilde{\gamma }_0 (w-\log (w)-1)\Big]\nonumber\\
  S_2(z)=&T_{2,0} w^{-\tilde{\gamma }_0-2}-T_{1,0} w^{-\tilde{\gamma }_0-2} \Big[(w-1) \tilde{\gamma }_1\nonumber\\&+\tilde{\beta }_1 \left(\tilde{\gamma }_0 (-w+\log (w)+1)+\log (w)\right)\Big]\nonumber\\&+\frac{1}{2} T_{0,0} w^{-\tilde{\gamma }_0-2} \Big\lbrace-\tilde{\beta }_1 \tilde{\gamma }_1 \Big[1-w^2+2 \log (w)\nonumber\\&+2 (w-1) \tilde{\gamma }_0 (w-\log (w)-1)\Big]+(w-1)\nonumber\\&\times \Big[(w-1) \tilde{\beta }_2 \tilde{\gamma }_0+(w-1) \tilde{\gamma }_1^2-(w+1) \tilde{\gamma }_2\Big]\nonumber\\&+\tilde{\beta }_1^2 \tilde{\gamma }_0\big(\tilde{\gamma }_0-1\big) (w-\log (w)-1)^{2} \Big\rbrace\,,
  \label{eq:rgsol}
 \end{align}
 where $w\equiv 1-\beta_0 \hs z$, for anomalous dimension and higher order beta function coefficients, we have used $\tilde{X}\equiv X/\beta_0$. The important feature of the above procedure is that the most general term of RGSPT is given by:
 \begin{align}
     \Omega_{n,a}\equiv\frac{\log^n(w)}{w^a}=\frac{\log^n(1-\beta_{0}\hs x(\mu)\log(\mu^2/Q^2))}{(1-\beta_{0}\hs x(\mu)\log(\mu^2/Q^2))^a}\,,
     \label{eq:coef_rgspt}
 \end{align}
 where $n$ is a positive integer and $a\propto \gamma_0/\beta_{0}$ appearing in Eq.~\eqref{eq:anom_S}. It should be noted that for $\mu^2=Q^2$, both RG summed series in Eq.~\eqref{eq:summed_ser} and Eq.~\eqref{eq:Pseris} agree with each other. The analytic continuation for them is obtained by taking discontinuity of $\log(\mu^2/Q^2)=\log(\mu^2/|Q|^2)\pm i \hs \pi$. This procedure results in large ``$i\hs \pi$" corrections for FOPT, but for RGSPT, such corrections are summed to all orders in the terms like in Eq.~\eqref{eq:coef_rgspt}. For numerical prescriptions such as CIPT, the analytic continuation is obtained by using Eq.~\eqref{eq:master_relation}. One important point to note here is that results from different prescriptions, such as RGSPT and FOPT, are not the same when $\mu^2=Q^2$ is set after operations like analytic continuation or Borel transformation are performed. The differences arise due to different treatments of the RG logarithms for finite order series for which only a few terms are known. For more details on analytic continuation using FOPT and RGSPT, we refer to Ref.~\cite{AlamKhan:2023dms}.\par
The reduced sensitivity on the renormalization scale in RGSPT prescription is due to the cancellation between running parameters (coupling and masses by numerically solving Eq~\eqref{eq:RGEx} and Eq.~\eqref{eq:RGE}) and coefficients $S_i(z)$ at different orders. For a simpler case, in which series Eq.~\eqref{eq:Pseris} with vanishing anomalous dimension ($\gamma_i=0$ in Eq.~\eqref{eq:rgsol}), there is a perfect cancellation between $S_1(z)$ in Eq.~\eqref{eq:rgsol} and exact one-loop running of the strong coupling constant ($x(\sqrt{\xi} \mu)=\frac{x(Q)}{1-x(Q) \beta_0 \log(Q^2/\left(\xi \mu^2\right))}+\order{x^2}$). It is not easier to such perfect cancellation for higher orders as the exact analytical solution to $x$, using Eq.~\eqref{eq:RGEx}, is not known.\par An alternate way to achieve RG improvement and access would be to rearrange the original series in Eq.~\eqref{eq:Pseris} by replacing running logarithms as:
  \begin{align}
      \log(\frac{\mu^2}{Q^2})=-\log(\xi)+ \log(\xi\frac{\mu^2}{Q^2})=-\log(\xi)+L_\xi\,,
  \end{align}
 and get a closed for summation, similar to Eq.~\eqref{eq:summed_ser}, that has form:
\begin{align}
    \mathcal{S}^{\Sigma_\xi}=\sum_i x(\mu)^i \hs S_i(\xi,x(\mu)\hs L_\xi)\,.
\end{align}
  Such a summation is not performed for RGSPT in the literature and is left for future studies\footnote{We thank the referee for pointing out this alternate method to explore the scale dependence in RGSPT.}.
\section{Hadronic spectral function}\label{sec:hadronic_info}
The hadronic spectral functions are constructed using the contributions from the pion/kaon pole and the data from the experiments on the resonances in a given channel. At low energies, they are dominated by the pion/kaon pole contributions. This section discusses the parametrization of the unknown pseudoscalar spectral function for the non-strange and strange channels. 
\subsection{Non-strange channel}\label{subsec:rho_NS}
\begin{widetext}
    
\begin{figure}[H]
\centering
	\begin{subfigure}{.46\textwidth}
		\includegraphics[width=\linewidth]{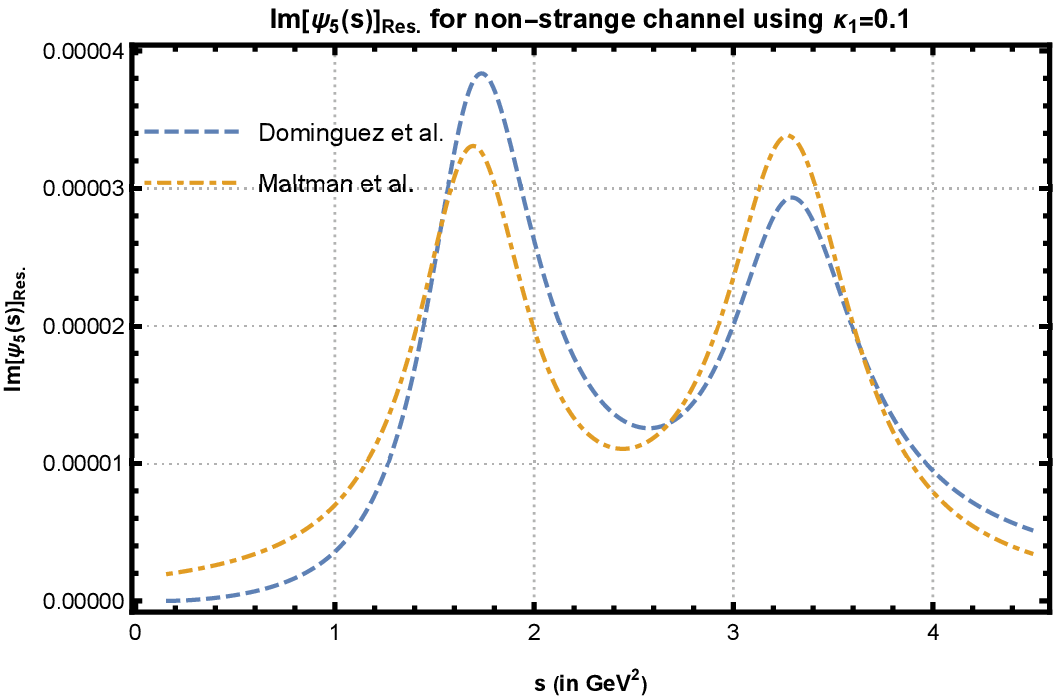}
	\end{subfigure}
	\begin{subfigure}{.45\textwidth}
		\includegraphics[width=\linewidth]{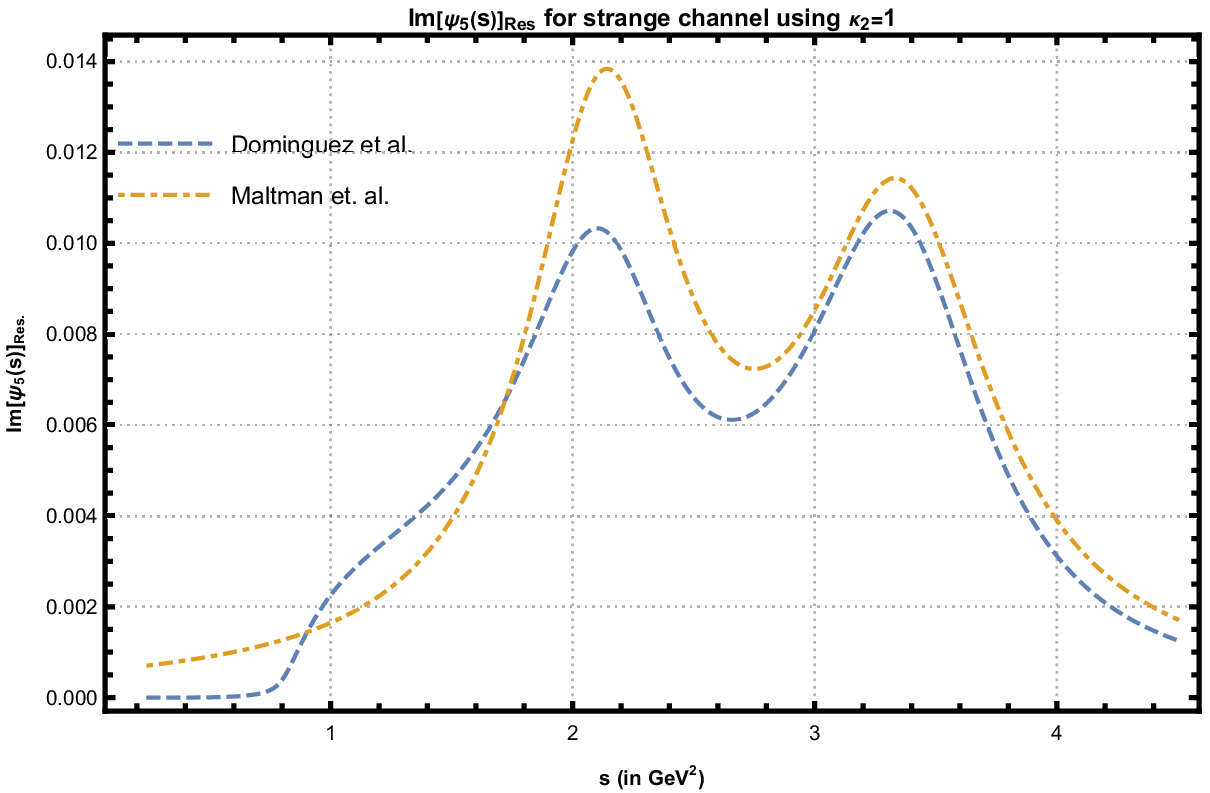}
 \end{subfigure}
\caption{\label{fig:had} Hadronic spectral function in the resonance region for strange and non-strange channels.}
\end{figure}

\end{widetext}

For the non-strange pseudoscalar channel, two phenomenological parametrizations are often used in the literature. In Ref.~\cite{Dominguez:1986aa}, Dominguez and Rafael provided a ChPT-based parametrization which is normalized to unity at the threshold. Later some corrections are reported for this parametrization in Ref.~\cite{Bijnens:1994ci}. Another parametrization often used is by Maltman and Kambor~\cite{Maltman:2001gc}, which requires masses and decay constants for the higher resonances. In this article, we have used Dominguez and Rafael's parametrization, which is recently used in Ref.~\cite{Dominguez:2018azt} for the up/down-quark mass determination. We have used their results for the hadronic parametrization, which is given by:
\begin{align}
    \rho_{\text{NS}}= f_\pi^2 M_\pi^4\hs \delta\left(s-M_\pi^2\right)+\rho_{3\pi} \frac{\text{BW}_1(s)+\kappa_1 \hs \text{BW}_2(s)}{1+\kappa_1}\,,
    \label{eq:had_NS}
\end{align}
where, $f_\pi=130.2(1.2)\hs\MeV$ and $M_\pi=130.2(1.2)\hs\MeV$ value of $\kappa_1\simeq0.1$ is used in the Ref.~\cite{Dominguez:2018azt} and it controls the relative importance of the resonances. The $3\pi$ resonance contributions are received from the $\pi(1300)$ and $\pi(1800)$ states. Their contributions are encoded in the $\rho_{3\pi}$, which is given by:
\begin{align}
\rho_{3\pi}&=\frac{1}{\pi}\hs\text{Im}\hs \Psi_{5}(s)\vert_{3\pi}\nonumber\\
&=\frac{1}{9}\frac{M_{\pi}^2}{f_{\pi}^2}\frac{1}{2^7\pi^4}\theta\left(s-9 M_{\pi}^2\right) I_{\pi}(s)\,,
\end{align}
where, $I_{\pi}(s)$ is the phase space integral given in Eq.~\eqref{eq:I_pi}. In the chiral limit, the phase integral reduces to $I_\pi(s)=3\hs s$ that confirms the prediction for $\rho_{3\pi}$ in Ref.~\cite{Pagels:1972xx}. The $\text{BW}_{1,2}(s)$ is the Breit-Wigner distribution given by:
\begin{align}
\text{BW}_i(s)=\frac{\left(M_i^2-s_{\text{th}}\right)^2+M_i^2\Gamma_i^2}{\left(s-M_i^2\right)^2+M_i^2\Gamma_i^2}
\end{align}
which is normalized to unity at the threshold, i.e., $\text{BW}_{1,2}(s_{th})=1$. For the non-strange spectral function, we use the following data from PDG~\cite{ParticleDataGroup:2022pth} as input:
\begin{align}
     M_\pi&=134.9768(5)\MeV, f_\pi=130.2(1.2)\MeV\nonumber\,,\\
    M_{\pi,1}&=1300(100)\MeV,\quad \Gamma_{\pi,1}=260(36)\MeV\nonumber\,,\\
    M_{\pi,2}&=1810^{+11}_{-9}\MeV,\quad \Gamma_{\pi,2}=215_{-8}^{+7}\MeV\nonumber\,.
\end{align}
Using the above values as input, the non-strange spectral function in the two parameterizations discussed above is plotted in Fig.~\eqref{fig:had}. 

\subsection{The Strange Channel}
We use the hadronic parametrization presented in Ref.~\cite{Dominguez:1997eu} for the strange channel. This parametrization is equivalent to the one we have used in the non-strange channel. The hadronic spectral function is given by:
\begin{align}
    \rho_S(s)&=\frac{1}{\pi}\text{Im}\hs \Psi_5(s)\vert_\text{Had.}\nonumber\\&=f_K^2 M_K^2\delta\left(s-M_K^2\right)+\frac{1}{\pi}\text{Im}\left(\Psi_5(s)\right)\vert_{\text{Res.}}\,.
    \label{eq:had_S}
\end{align}
where the spectral function for the resonance region is given by 
\begin{align}
 \frac{1}{\pi}\text{Im}\left(\Psi_5(s)\right)\vert_{\text{Res.}}=
 \rho_{K\pi\pi}(s)\frac{\text{BW}_1(s)+\kappa_2 \hs \text{BW}_2(s)}{1+\kappa_2}
\end{align}
The Breit-Wigner profile is constructed from $K(1460)$ and $K(1830)$ resonances. The value $\kappa_2\simeq1$ is found to be a reasonable choice in Ref.~\cite{Dominguez:1997eu} to control the contributions from the resonances. In addition, due to its narrow width, there is a significant contribution from the resonant sub-channel $K^*(892)-\pi$. Its contributions are  also included in the $\rho_{K\pi\pi}(s)$ and has the form:
\begin{align}
    \rho_{K\pi\pi}(s)=\frac{M_K^2}{2f_\pi^2}\frac{3}{2^7\pi^4}\theta(s-M_K^2)\frac{I_K(s)}{s\left(M_K^2-s\right)}\,,
\end{align}
and the integral $I_K(s)$ is defined in Eq.~\eqref{eq:Is}. For the strange channel, precise values of the resonance masses and the decay width do not exist. We are using the values used in Ref.~\cite{Schilcher:2013pvu} with additional uncertainties of $50 \MeV$ to resonance masses and $10\%$ to the decay widths. For kaons, we use PDG~\cite{ParticleDataGroup:2022pth} values, and the following values for the parameters for the strange channel are used as input:
\begin{align}
    M_K&=497.611(13)\MeV, \quad f_K=155.7(3)\MeV\nonumber\,,\\
    M_{K,1}&=1460(50)\MeV,\quad \Gamma_1=260(26)\MeV\nonumber\,,\\
    M_{K,2}&=1830(50)\MeV,\quad \Gamma_1=250(25)\MeV\nonumber\,.\\
    M_{K^*}&=895.55(2)\MeV,\quad \Gamma_{K^*}=47.3(5)\MeV\,.
\end{align}
Using the above inputs, the strange spectral function in the resonance region is plotted in Fig.~\eqref{fig:had}.

\begin{widetext}

\begin{align}
I_\pi(s)=\int_{4 M_\pi^2}^{\left(\sqrt{s}-M_\pi\right)^2} &  d\hs u\hs \sqrt{1-\frac{4 M_\pi^2}{u}}\frac{\lambda^{1/2}(s,u,M_\pi^2)}{s}\Bigg\lbrace 5+\frac{1}{\left(s-M_\pi^2\right)}\left[3\left(u-M_\pi^2\right)-s+9M_\pi^2\right]\nonumber\\&+\frac{1}{2\left(s-4 M_\pi^2\right)^2}\bigg[\left(s-3u+3 M_\pi^2\right)^2+3\lambda(s,u,M_\pi^2)\left(1-4\frac{M_\pi^2}{u}\right)+20M_\pi^4 \bigg]\Bigg\rbrace
\label{eq:I_pi}
\end{align}
where, $\lambda(s,u,M_\pi^2)$ is given by:
\begin{align}
    \lambda(s,u,M_\pi^2)=\left(s-\left(\sqrt{u}-M_\pi\right)^2\right)\left(s-\left(\sqrt{u}+M_\pi\right)^2\right)\,,
\end{align}
 
\begin{align}
    I_K(s)=\int_{M_K^2}^{s}\frac{d\hs u}{u}\left(u-M_K^2\right)\left(s-u\right)\Bigg \lbrace \left(M_K^2-s\right)\left[u-\frac{(s+M_K^2)}{2}     \right]-\frac{1}{8u}\left(u^2-M_K^4\right)\left(s-u\right) +\frac{3}{4} \left(u-M_K^2\right)^2\vert F_{K^*}(u)\vert^2 \Bigg \rbrace
    \label{eq:Is}
\end{align}
and
\begin{align}
    \vert F_{K^*}(u)\vert^2=\frac{\left(M_{K^*}^2-M_K^2\right)^2+M_{K^*}^2\Gamma_{K^*}^2}{\left(M_{K^*}^2-u\right)^2+M_{K^*}^2\Gamma_{K^*}^2}
\end{align}

\end{widetext}

\section{The OPE corrections}
\label{sec:OPE}
The OPE corrections are calculated in large $Q^2$ limit, and organized as expansion in $1/Q^2$:
\begin{align}
	\Psi_{5}(Q^2)=Q^2 \hs  \left(m_1+ m_2\right)^2 \sum_{i=0}\frac{\Psi_{2\hs i}(Q^2)}{(Q^2)^i}\,, 
 \label{eq:corr_exp}
\end{align}
 and $\Psi_{2\hs n}(Q^2)$ are termed as the contributions from $2n$-dimensional operators for $n=0,1,2,3,\cdots$. Quantities $\Psi_0(Q^2)$ and $\Psi_2(Q^2)$ are purely perturbative while additional non-perturbative condensate corrections start from $\Psi_4(Q^2)$. The leading term in the OPE, 
$\Psi_0(Q^2)$, is known to $\ordas{4}$
 \cite{Gorishnii:1990zu,Chetyrkin:1996sr,Baikov:2005rw} which has an expansion in terms of $\as$ and $\log(\mu^{2}/Q^{2})$. The dimension-2 term of OPE, $\Psi_2(Q^2)$, receives massive corrections $(\propto m_i^2)$ and it is known to $\ordas{1}$~\cite{Generalis:1990id,Chetyrkin:1985kn,Generalis:1990id,Jamin:1992se,Jamin:1994vr}. Additional strange quark mass corrections ($\propto \order{m_s^2\as^2}$) to it are included from Ref.~\cite{Chetyrkin:2005kn}. The non-logarithmic terms appearing in $\Psi_0(Q^2)$ and $\Psi_2(Q^2)$ are irrelevant for the Sum rule in Eq.~\eqref{eq:bs_final}. This is due to the fact that the contributions calculated from the Borel operator, in Eq.~\eqref{eq:BO}, vanish for any non-negative integer powers of $Q^2$. For the spectral function case, non-logarithmic terms vanish when analytic continuation is performed using Eq.~\eqref{eq:spectral_density}. The relevant expressions for these quantities for FOPT and RGSPT can be found in appendix~\eqref{app:dim0adler} and \eqref{app:dim2adler}. 

The OPE contributions from the dimension-4 term, $\Psi_4(Q^2)$, contain both massive corrections~($\propto\hs m_i^4$) as well as non-perturbative condensates of quarks and gluon fields. These corrections are known to $\ordas{1}$~\cite{Pascual:1981jr,Jamin:1992se,Jamin:1994vr}. Their RG running should also be taken into account when coupling and masses are evolved with the scale. However, we use the results provided in Refs.~\cite{Spiridonov:1988md,Baikov:2018nzi} to form an RG invariant combination of these condensates. For the quark condensates, this relation is given by:
\begin{align}
	\langle m_i \overline{q}_j\hs q_j\rangle_{\text{inv.}}=\langle m_i \overline{q}_j \hs q_j \rangle +m_i\hs  m_j^3\left(\frac{3}{7 \pi^2\hs x }-\frac{53}{56\pi^2}\right)\,.
\end{align}
The RG invariant combinations of the condensates \cite{Spiridonov:1988md} also introduce inverse powers of the $\as$~\cite{Jamin:1994vr,Chetyrkin:1994qu}. For the gluon condensate, we use the following relation:
\begin{align}
    \frac{\beta(x)}{x^2}\langle\frac{\as}{\pi}G^2\rangle_{\text{inv.}}\equiv&\frac{\beta(x)}{x^2}\langle\frac{\as}{\pi}G^2\rangle-4\gamma_m(x)\sum_{k={u,d,s}}\langle m_i \overline{q}_i\hs q_i\rangle\nonumber\\&-\frac{3}{4\pi^2}\gamma_{\text{vac.}}\sum_{k={u,d,s}} m_k^4
\end{align}
where, $\gamma_{\text{vac.}}=-1 - (4 x)/3 + x^2 (-223/72 + 2/3 \zeta(3))$ is the vacuum anomalous dimension~\cite{Baikov:2018nzi}. The expression for $\Psi_4$ in RGSPT and FOPT are provided in the appendix~\eqref{app:dim4}.\par

We also consider the dimension-6 contribution to the OPE, $\Psi_{6}(Q^2)$, for which only condensate terms are known. These corrections can be written as:
\begin{align}
\Psi_{6}(Q^2)=\frac{\left(I_6\right)_{12}}{6}\,,\nonumber
\end{align}
where
\begin{align}
	\left(I_6\right)_{ij}=&-3\hs (m_i\langle  \overline{q}_{j}q_{j}\hs G\rangle+m_j\langle \overline{q}_{i}q_{i}\hs G\rangle)\nonumber\\&-\frac{32}{9}\pi^2 x \left(\langle\overline{q}_{i}q_{i}\rangle^2+\langle\overline{q}_{j}q_{j}\rangle^2-9\langle\overline{q}_{i}q_{i}\rangle\langle\overline{q}_{j}q_{j}\rangle \right)\,.
	\label{eq:dim6}
\end{align}
Subscript $i$ and $j$ stand for the quark flavors in the strange and non-strange channels. It should be noted that the structure of the dimension-6 condensate is rather complicated, and in deriving Eq.~\eqref{eq:dim6}, vacuum saturation approximation is used to relate dimension-6 four-quark condensate terms to dimension-4 quark condensates. For more details, we refer to~\cite{Shifman:1978bx,Jamin:1994vr}.\par 
The numerical values used for the non-perturbative quantities are as follows:
\begin{align}
    \langle \overline{u}u\rangle&=-\frac{f_\pi^2 M_\pi^2}{2\left(m_u+m_d\right)}\text{~\cite{Gell-Mann:1968hlm}}\,,\\
    \langle \overline{s}s\rangle&=\left(0.8\pm0.3\right) \langle \overline{s}s\rangle\text{~\cite{Chetyrkin:2005kn}}\,,\\
    \langle\frac{\as}{\pi}G^2\rangle&=0.037\pm0.015\GeV^4\text{~\cite{Dominguez:2014pga}}\,,\\
    \langle \overline{q}_{i}q_{i}\hs G\rangle&=M_0^2 \langle 
    \overline{q}_{i}q_{i}\rangle\text{\cite{Ioffe:2005ym}}\,,\\
M_0^2&=0.8\pm0.2\hs \GeV^2 \text{~\cite{Ioffe:2005ym}}\,.
\end{align}
We neglect the contributions to OPE beyond this order. \par
 From Eq.\eqref{eq:corr_exp}, we can obtain $\Psi''_{5}(Q^2)$ which has following form:
 \begin{align}
 		\Psi''_{5}(Q^2)=\frac{\left(m_1+ m_2\right)^2} {Q^2}\sum_{i=0}\frac{\tilde{\Psi}''_i(Q^2)}{(Q^2)^i}\,,
   \label{eq:PiD2}
 \end{align}
 and the Borel transform as:
 \begin{align}
     \Psi''_{5}(u)=\frac{\left(m_1+ m_2\right)^2} {u}\sum_{i=0}\frac{\tilde{\Psi}''_i(u)}{u^i}\,.
     \label{eq:psi_Borel}
 \end{align}
 The spectral function from Eq.~\eqref{eq:corr_exp} is obtained by using Eq.~\eqref{eq:spectral_density}, and it can be organized as:
\begin{align}
	\rho^{OPE}_{5}(s)&= s \hs \mathcal{R}_0(s)+\mathcal{R}_2(s)+\frac{1}{s}\mathcal{R}_4(s)+\frac{1}{s^2}\mathcal{R}_6(s)+\cdots\,,
	\label{eq:rho_exp}
\end{align}
where $\mathcal{R}_{n}$ are calculated from $\Psi_{n}$ using ~\eqref{eq:master_relation} and analytical expressions for $\mathcal{R}_0$ can be found in Ref.~\cite{Chetyrkin:2005kn}. It should be noted that $\rho^{OPE}_{5}(s)$ and $\Psi''_{5}(Q^2)$ are RG invariant perturbative quantities that enter in Borel-Laplace sum rule in Eq.~\eqref{eq:bs_final}.\par
 The  $\Psi_n(q^2)$ for the pseudoscalar current appearing in the OPE expansion of Eq.~\eqref{eq:corr_exp}, are not RG invariant quantity. The Adler function, $\mathcal{D}_{n}(Q^2)$, is obtained from it using the relation:
\begin{align}
    \mathcal{D}_{n}(Q^2)\equiv -Q^2 \frac{d}{d\hs Q^2}\left[\left(m_1+ m_2\right)^2\hs \Psi_n(Q^2)\right]\,,
    \label{eq:def_Adler}
\end{align}
which is RG invariant and $m_i=m_i(Q)$ as mentioned before in the text below Eq.~\eqref{eq:Psi2Pi}. Both $\Psi_{n}(Q^{2})$ and $\mathcal{D}_{n}(Q^{2})$ have a cut  $Q^2=-q^{2}<0$ due to the term $\log(\frac{\mu^{2}}{-q^{2}})$. The spectral density in Eq.~\eqref{eq:rho_exp} is obtained in the timelike regions ($s=q^2>0$) from the discontinuity of the polarization function:
	\begin{align}
		\mathcal{R}_{n}(s)&\equiv\frac{1}{2\pi i }\lim_{\epsilon\rightarrow 0}\left[ \Psi_n(-s-i \epsilon)-\Psi_n(-s+i \epsilon)\right]\nonumber\\&=\frac{1}{2\pi i}\int_{-s+i \epsilon}^{-s-i \epsilon} d\hs q^2 \frac{d}{d\hs q^2}\Psi_n(q^2)\nonumber\\&=\frac{-1}{2\pi i} \int_{-s+i \epsilon}^{-s-i \epsilon} \frac{dq^2}{q^2} \mathcal{D}_{n}(q^2)\nonumber\\ &=\frac{-1}{2\pi i}\oint_{|x_c|=1} \frac{d x_c}{x_c} \mathcal{D}_{n}(-x_c s)\,.
  \label{eq:master_relation}	
 \end{align}
The contour integral in the above equation has to be evaluated without crossing the cut for $q ^{2}>0$. It should be noted that for FOPT and RGSPT prescriptions, the imaginary part can be obtained trivially by replacing the $\log(\mu^2/Q^{2})=\log(\mu^2/|Q|^{2})\pm i \hs \pi$ across the cut. For the numerical evaluation methods, such as in the CIPT prescriptions, Eq.~\eqref{eq:master_relation} can be very useful for analytic continuation in the complex plane. To sum $\pi^{2}$-terms in RGSPT, we first perform the RG improvement of the $\Psi_n(q^{2})$ or $\mathcal{D}_n(q^{2})$. The resulting perturbative has the most general term given in Eq.~\eqref{eq:coef_rgspt} for which the imaginary part can be taken by simply setting  $\log(\mu^2/Q^{2})=\log(\mu^2/|Q|^{2})\pm i \hs \pi$. This process results in an analytic expression for which the renormalization scale can be set $\mu^{2}=s$, but the $i\pi$ terms are left behind, which results in improved convergence.  For more details about their effects on the summation of kinematical terms, we refer to \cite{AlamKhan:2023dms}. \par
It should be noted that the analytic continuation using the RGSPT expressions for the $\mathcal{R}_n$ are rather lengthy. Therefore, we provide expressions for corresponding Adler functions in appendix~\eqref{app:dim0adler}.
 \subsection{Analytic continuation in FOPT and RGSPT }\label{subsec:ancont}
\begin{figure}[ht]
    \centering
		\includegraphics[width=\linewidth]{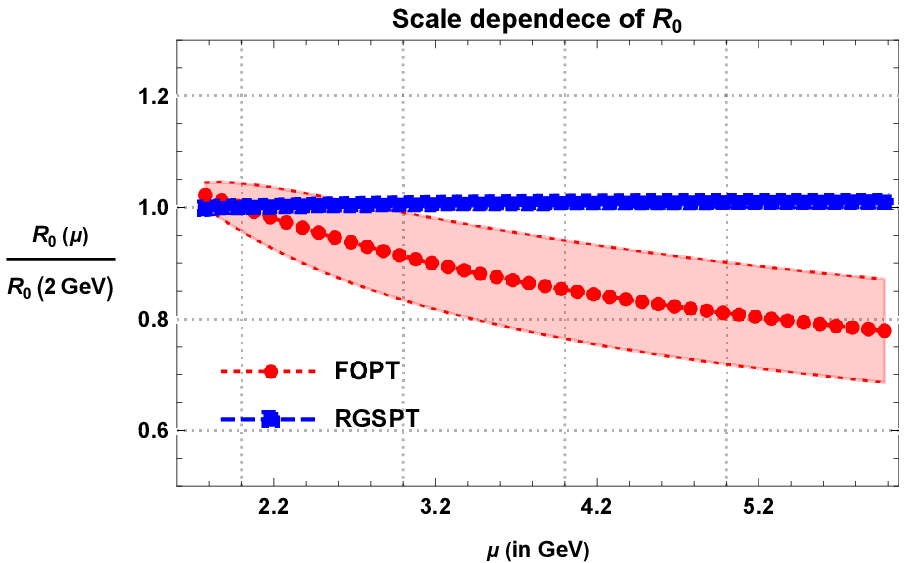}
  \caption{\label{fig:scdepR0}}
		
		\includegraphics[width=\linewidth]{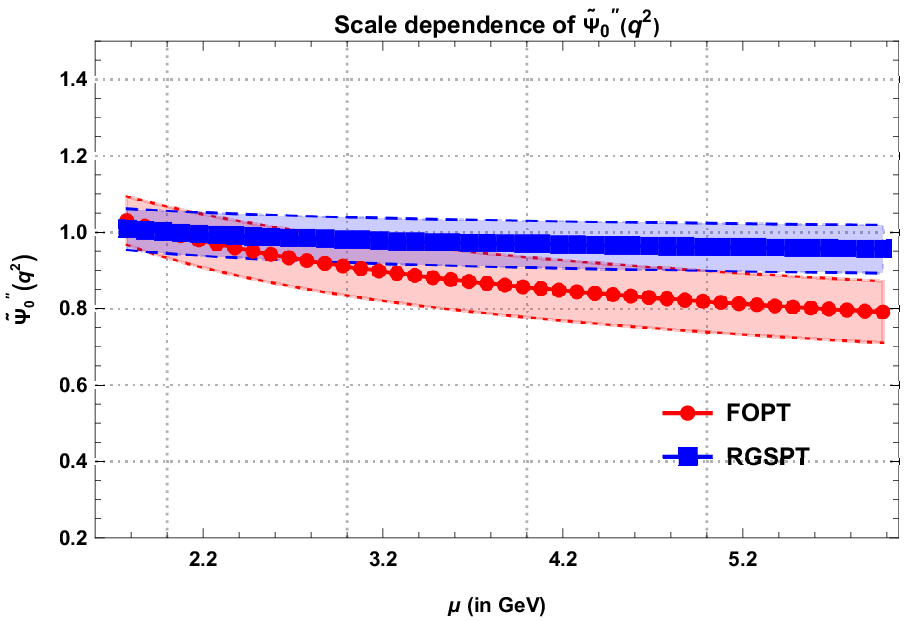}
 \caption{ \label{fig:scdeppsi0}}
    
    \caption{Renormalization scale dependence of $R_0(s)$ and $\overline{\Psi}''_0(q^2)$ normalized to unity at $2\hs\GeV$ in RGSPT and FOPT. The bands represent the truncation uncertainty.}
    \label{fig:scdepR0Pd2}
\end{figure}

The $\mathcal{R}_i(s)$ are obtained from $\Psi_n(q^2)$ by its analytic continuation from spacelike regions to timelike regions (using Eq.~\eqref{eq:master_relation}), which results in the large kinematical $\pi^2$ corrections. These corrections, however, can be summed to all orders using RGSPT, and a good convergence is obtained for the perturbative series. As a demonstration, we define $R_0$ from $\mathcal{R}_0$ as:
\begin{align}
    R_0\equiv\frac{8\pi^2 }{3 (m_s(2))^2}\mathcal{R}_0\,.
    \label{eq:R0temp}
\end{align}
Using $\as(2\GeV)=0.2945$ and $m_s(2\GeV)=93.4\MeV$ and setting $m_u=0$, the $R_0$ at different orders of $\as$ has the following contributions:
\begin{align}
    R_0^{\text{FOPT}}&=1.0000+0.6612+ 0.4909+ 0.2912+ 0.1105\,,\\
    R_0^{\text{RGSPT}}&=1.0038+0.4175+0.1760+ 0.0581 -0.0152\,.
\end{align}
We can see that summation of the $\pi^2$-terms enhances the convergence of the perturbation series when RGSPT is employed. The scale dependence of the $R_0$ and truncation uncertainty at different scales are significantly improved, which can be seen in Fig.~\eqref{fig:scdepR0}. \par
We can also test the RG improvement for the $\tilde{\Psi}_0''(q^2)$ by defining $\overline{\Psi}''_0(q^2)$, analogous to Eq.~\eqref{eq:R0temp}, as:
\begin{align}
     \overline{\Psi}''_0(q^2)\equiv\frac{8\pi^2 }{3 (m_s(2\GeV))^2}\tilde{\Psi}_0\,.
    \label{eq:psi0temp}
\end{align}
and using quark masses and $\as$ same as $R_0$ and setting $q=2\GeV$, we get the following contributions to $\overline{\Psi}^{'',\text{FOPT}}_0(q^2)$:
\begin{align}
    \overline{\Psi}^{'',\text{FOPT}}_0&=1.0000+0.4737+ 0.2837+ 0.1917+ 0.1405\,,\\
  \overline{\Psi}^{'',\text{RGSPT}}_0&=1.1508+ 0.5280+ 0.2621+ 0.1670+ 0.1244\,.
\end{align}
In the case of $\overline{\Psi}''_0(q^2)$, we get slightly better convergence than FOPT. The scale dependence of $\overline{\Psi}''_0(q^2)$ normalized to unity at $2\hs\GeV$ is plotted in Fig.~\eqref{fig:scdeppsi0}.

\subsection{Borel Transform in FOPT and RGSPT} \label{subsec:BorelTransform}
The perturbative series in the FOPT prescription is a polynomial form containing $\as$, $m^2_q/Q^2$, and $\log(\frac{\mu^2}{Q^2})$. For Borel transforms in FOPT, only terms $\log(\frac{\mu^2}{Q^2})$ and $Q^2$ are relevant, and we get an analytical expression containing Euler's constant, Zeta functions in addition to terms $\log(\frac{\mu^2}{u})$ and powers of $u$. To obtain the Borel transform, we use the relation from Ref.~\cite{Jamin:1994vr} for the operator in Eq.~\eqref{eq:BO}:
\begin{align}
    \hat{\mathcal{B}}_{u}\Big[\frac{1}{(Q^2)^\alpha}&\log^n\left(\frac{\mu^2}{Q^2}\right)\Big]\nonumber\\=&\frac{1}{(u)^\alpha}\sum_{k=0}^n (-1)^k \hs\hs\text{}^n C_k\log^k\left(\frac{\mu^2}{u}\right)\del_\alpha^{n-k}\left(\frac{1}{\Gamma[\alpha]}\right)\,,
    \label{eq:B_FOPT}
\end{align}
where, $\text{}^n C_k=\frac{n!}{k!(n-k)!}$ is the binomial coefficient. The derivative of the Gamma function results in the appearance of the Euler's constant and Zeta functions that can be seen in Eq.~\eqref{eq:Borel_gammazeta}.\par
For RGSPT, Borel transform is not a trivial task; it involves transcendental function as encountered in Ref.~\cite{deRafael:1981bs} and is evaluated numerically. The most general term in RGSPT, from Eq.~\eqref{eq:coef_rgspt}, can be written as:
\begin{equation}
    \frac{\log^n(w)}{w^\alpha}= \left[\del_\delta^n\hs w^{\delta-\alpha}\right]_{\delta\rightarrow 0}\,,
\end{equation}
where $w=1-\beta_{0} x \log(\mu^2/s)$ and $\alpha$ is some real number depending upon the anomalous dimension of the quantity under consideration. \par
Using Schwinger parametrization, we can write:
\begin{equation}
\begin{aligned}
\frac{1}{w^{\alpha}}=&\frac{1}{\Gamma[\alpha]}\int_{0}^{\infty}dt\hs\hs t^{\alpha-1}\hs e^{-t\hs w}\\
=&\frac{1}{\Gamma[\alpha]}\int_{0}^{\infty}dt\hs\hs t^{\alpha-1}\hs e^{-t\hs \left(1-\beta_0 x \log(\mu^2/s)\right)}\\
=&\frac{1}{\Gamma[\alpha]}\int_{0}^{\infty}dt\hs\hs t^{\alpha-1}\hs \left(\mu^2/s\right)^{\beta_0 \hs x \hs t}\hs e^{-t}\\
=&\frac{1}{\Gamma[\alpha]}\sum_{n=0}^{\infty}\frac{(-1)^n}{\Gamma[n+1]}\int_{0}^{\infty}dt\hs\hs t^{\alpha+n-1}\hs \left(\mu^2/s\right)^{\beta_0 \hs x \hs t}
\end{aligned}
\label{eq:Sch_par}
\end{equation}
Using the above relation, we can easily perform the Borel operator as follows:
	\begin{align}
		\hat{\mathcal{B}}_{u}\left[\frac{1}{s^{z}}\frac{1}{w^{\alpha}}\right]=&\frac{1}{(\mu^2)^z\Gamma[\alpha]}\sum_{n=0}^{\infty}\frac{(-1)^n}{\Gamma[n+1]}\nonumber\\&\bs\times\int_{0}^{\infty}dt\hs \frac{t^{\alpha+n-1}}{\Gamma[z+\beta_0 x t]}\hs \left(\mu^2/u\right)^{\beta_0 \hs x \hs t+z}\,,
  \label{eq:BO_RGtemp}
	\end{align}
	where we have used the identity:
	\begin{align}
		\hat{\mathcal{B}}_{u}\left[\frac{1}{s^{\alpha}}\right]=\frac{1}{\Gamma[\alpha]u^{\alpha}}\,.
	\end{align}
	Now, we re-scale integral in Eq.~\eqref{eq:BO_RGtemp} by substituting $\tilde{t}=\beta_0\hs  x\hs t$ and rewrite it as:
	\begin{align}
	\hat{\mathcal{B}}_{u}\left[\frac{1}{s^{z}}\frac{1}{w^{\alpha}}\right]=&\frac{1}{(\mu^2)^z\Gamma[\alpha]}\sum_{n=0}^{\infty}\frac{(-1)^n}{\Gamma[n+1] \left(\beta_0 \hs x\right)^{n+\alpha}}\nonumber\\&\bs\int_{0}^{\infty}d\tilde{t}\hs \frac{\tilde{t}^{\alpha+n-1}}{\Gamma[z+\tilde{t}]}\hs \left(\mu^2/u\right)^{\tilde{t}+z}\,.
	\label{eq:bs1}
	\end{align}
 We can see that integral in the above relation can not be evaluated analytically~\cite{bateman1981higher}. We use:
	\begin{align}
	\tilde{\mu}(z,b,a)\equiv\int_{0}^{\infty}dt\frac{x^{a+t}t^{b}}{\Gamma[b+1]\Gamma[a+t+1]}
	\label{eq:identity1}
\end{align}
 to rewrite Eq.~\eqref{eq:bs1} as:
\begin{align}
	\hat{\mathcal{B}}_{u}\left[\frac{1}{s^{z}}\frac{1}{w^{\alpha}}\right]=&\frac{1}{(\mu^2)^z\Gamma[\alpha]}\sum_{n=0}^{\infty}\frac{(-1)^n \Gamma[\alpha+n-1]}{\Gamma[n+1] \left(\beta_0 \hs x\right)^{n+\alpha}}\nonumber\\&\bs\times\tilde{\mu}(\mu^2/u,\alpha+n-1,z)\,.
	\end{align}
We have to rely on numerical methods beyond this point. However, the identity:
\begin{align}
	\int_{0}^{\infty}e^{-s t} \tilde{\mu}(t,b,a) dt=s^{-\alpha-1}(\log(s))^{-\beta-1}\,,
\end{align}
allows us to recover the original function using Laplace transform.\par
Now, we can demonstrate the impact of the resummation for the Borel transformation. Consider leading mass corrections at different dimension to $\tilde{\Psi}''_j(s)$ from RGSPT, which has the following form:
\begin{align}
   A_j^{\text{RGSPT}}= \frac{1}{s(1-\beta_0 x L)^{(2j+2)\gamma_0/\beta_0}}\,,
\end{align}
where $L=\log(\mu^{2}/s)$ is used here for the discussion. Its series expansion to $\ordas{4}$ in FOPT is given by:
\begin{align}
   A_j^{\text{FOPT}}= &\frac{1}{s}\bigg(1+2 \gamma _0 L (j+1) x\nonumber\\&+\gamma _0 L^2 (j+1) x^2 \left(\beta _0+2 \gamma _0(1+j)\right)\nonumber\\&+\frac{2}{3} \gamma _0 L^3 (j+1) x^3 \left(\beta _0+(1+j)\gamma _0 \right)\nonumber\\&\bs\times \left(\beta _0+2 \gamma _0(1+j)\right)\nonumber\\&+\frac{1}{6} \gamma _0 L^4 (j+1) x^4\left(\beta _0+\gamma _0(1+ j)\right)\nonumber\\&\bs\times \left(\beta _0+2 \gamma _0(1+j) \right) \left(3 \beta _0+2 \gamma _0(1+j)\right)\bigg)\nonumber\\&+\ordas{5}\,.
   \label{eq:Bexp}
\end{align}
Now, the Borel transformation of the above series can be obtained by substituting the following values:
\begin{align}
	\label{eq:Borel_gammazeta}
\hat{\mathcal{B}}_u\left[\frac{1}{s}\right]&= \frac{1}{u}\,,\quad \hat{\mathcal{B}}_u\left[\frac{\log \left(\frac{\mu ^2}{s}\right)}{s}\right]= \frac{\log(\frac{\mu^2}{u})+\gamma_E }{u}\,,\nonumber\\\hat{\mathcal{B}}_u\left[\frac{\log ^2\left(\frac{\mu ^2}{s}\right)}{s}\right]&= \frac{\log^2(\frac{\mu^2}{u})+2 \gamma_E  \log(\frac{\mu^2}{u})+\gamma_E ^2-\zeta(2)}{u}\,,\nonumber\\\hat{\mathcal{B}}_u\left[\frac{\log ^3\left(\frac{\mu ^2}{s}\right)}{s}\right]&=\frac{\left(3 \gamma_E ^2-3\zeta(2)\right) \log(\frac{\mu^2}{u})+3 \gamma_E  \log^2(\frac{\mu^2}{u})}{u}\nonumber\\&\hs+\frac{\log^3(\frac{\mu^2}{u}) +2 \zeta (3)+\gamma_E ^3-3\gamma_E \zeta(2)}{u}\,,\nonumber\\\hat{\mathcal{B}}_u\left[\frac{\log ^4\left(\frac{\mu ^2}{s}\right)}{s}\right]&=\frac{8 \gamma_E  \zeta (3)+\gamma_E ^4+3/2\zeta(4)-6\gamma_E ^2 \zeta(2)}{u}\nonumber\\&+\frac{\log(\frac{\mu^2}{u}) \left(8 \zeta (3)+4 \gamma_E ^3-12 \gamma_E \zeta(2)\right)}{u}\nonumber\\&+\frac{\left(6 \gamma_E ^2-6\zeta(2)\right) \log^2(\frac{\mu^2}{u})+ \log^4(\frac{\mu^2}{u})}{u}\nonumber\\&+\frac{4 \gamma_E  \log^3(\frac{\mu^2}{u})}{u}%
\end{align}
where, $\gamma_E$ is Euler's constant, $\zeta(i)$ are the Zeta functions. These induced terms as a property of Borel-Laplace sum rules are first pointed in Ref.~\cite{Narison:1981ts} \footnote{We thank Prof. Narison for bringing this reference to our attention.}. It is interesting to note that all the $\gamma_E$ can be absorbed in the logarithms i.e. $\log(\frac{\mu^2e^{\gamma_E}}{u})$ but not the Zeta functions. A similar case for the Fourier transform of the static potential from momentum to the position space can be found in Ref.~\cite{Jezabek:1998wk}. \par
\begin{figure}[H]
\centering
			\includegraphics[width=\linewidth]{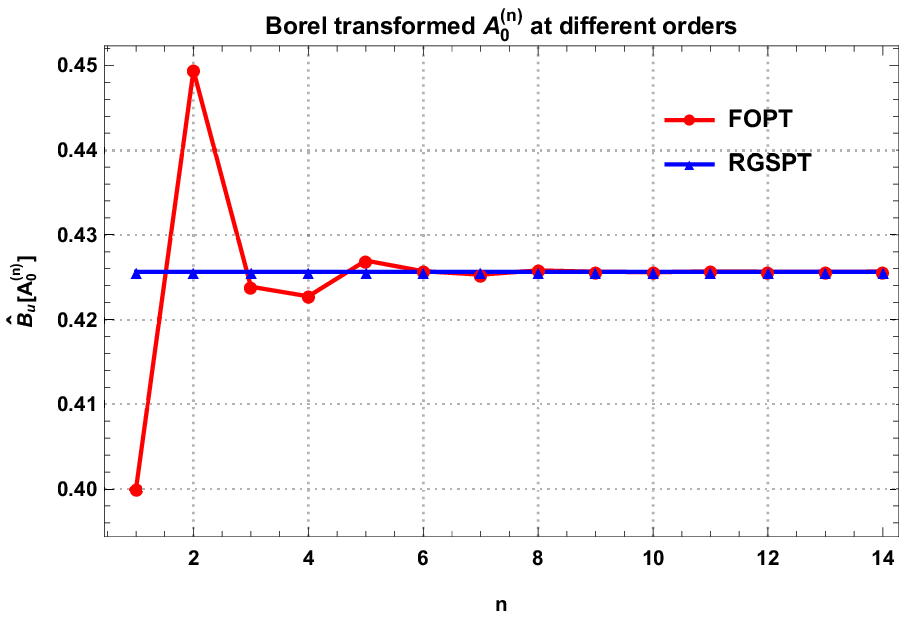}
		\includegraphics[width=\linewidth]{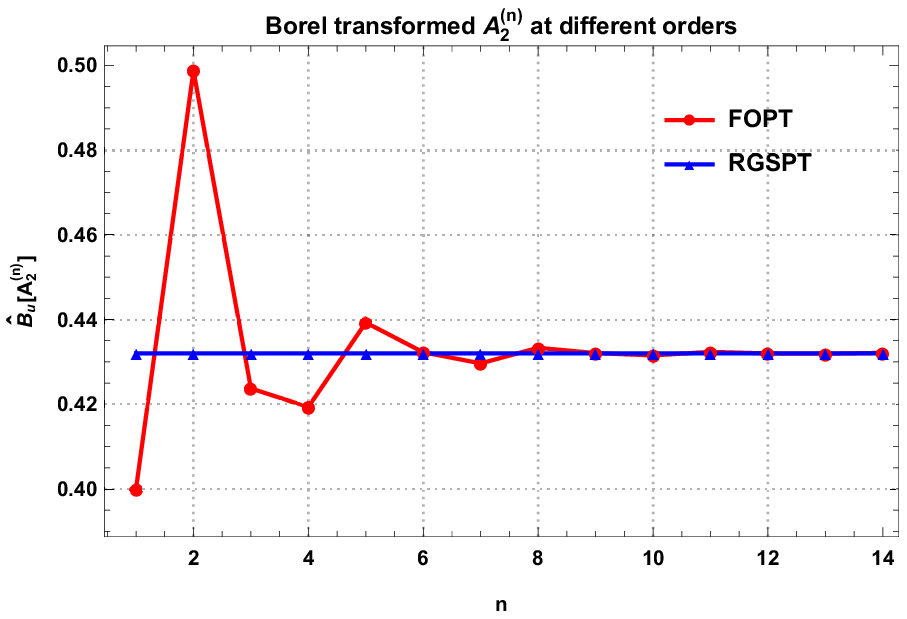}
	\includegraphics[width=\linewidth]{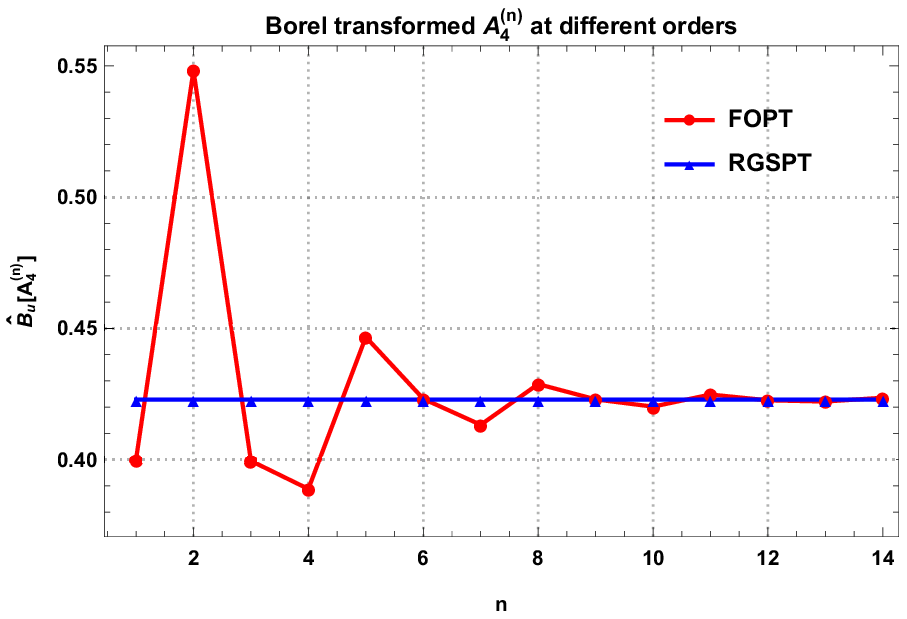}
 \caption{$A_n$ calculated at different orders using $\mu^2=u=2.5\GeV^2$. }
 \label{fig:A0}
    \end{figure}

Now, we obtain the Borel transform for $A_0$ using Eq.~\eqref{eq:Borel_gammazeta} and by setting $\mu^2=u=2.5\GeV^2$ that resums the logarithms in the case of FOPT. Using $x(\sqrt{2.5})=\as(\sqrt{2.5})/\pi=0.3361/\pi$, the Borel transformation of $A_0$ has the following contributions:
\begin{align}
    \hat{\mathcal{B}}_u [A_0^{\text{RGSPT}}]&=0.4256\,,\nonumber\\
     \hat{\mathcal{B}}_u [A_0^{\text{FOPT}}]&=0.4000+ 0.0494-0.0255 -0.0011+ 0.0042\nonumber\\&=0.4270\,.
\end{align}
It is clear from these numerical contributions that numerical contributions from leading logarithms are oscillatory, and the Borel transformed has poor convergence. These oscillations are due the the non-logarithmic terms of Eq.~\eqref{eq:Borel_gammazeta} (as we have set $\log(\frac{\mu^2}{u})=0$). The convergence gets worse for higher $j$ values, which can be inferred from Eq.~\eqref{eq:Bexp}, and the first three are plotted in Figs.~\eqref{fig:A0}. The RGSPT value is all order results, but for FOPT, it oscillates and slowly converges to the RGSPT value.\par
\begin{figure}[H]
    \centering
    \includegraphics[width=\linewidth]{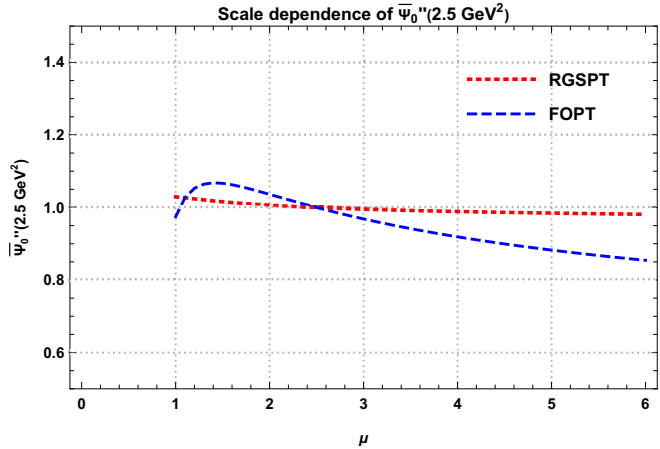}
    \caption{Scale dependence of the $\tilde{\Psi}''_0(u)\vert_{u=2.5\GeV^2}$ in RGSPT and FOPT.}
    \label{fig:RGP2u}
\end{figure}

 
We can use the above results to study the renormalization scale dependence of $\tilde{\Psi}''_0(u)$. To compare FOPT and RGSPT results, we use values of $\tilde{\Psi}''_0(u)\vert_{u=2.5\GeV^2}$ in these prescriptions, normalized to unity at $\mu=2.5\GeV$, and present our results in Fig.~\eqref{fig:RGP2u}. Again, results for RGSPT are very stable for a wide range of renormalization scales.

\begin{figure}[ht]
\centering
		\includegraphics[width=.48\textwidth]{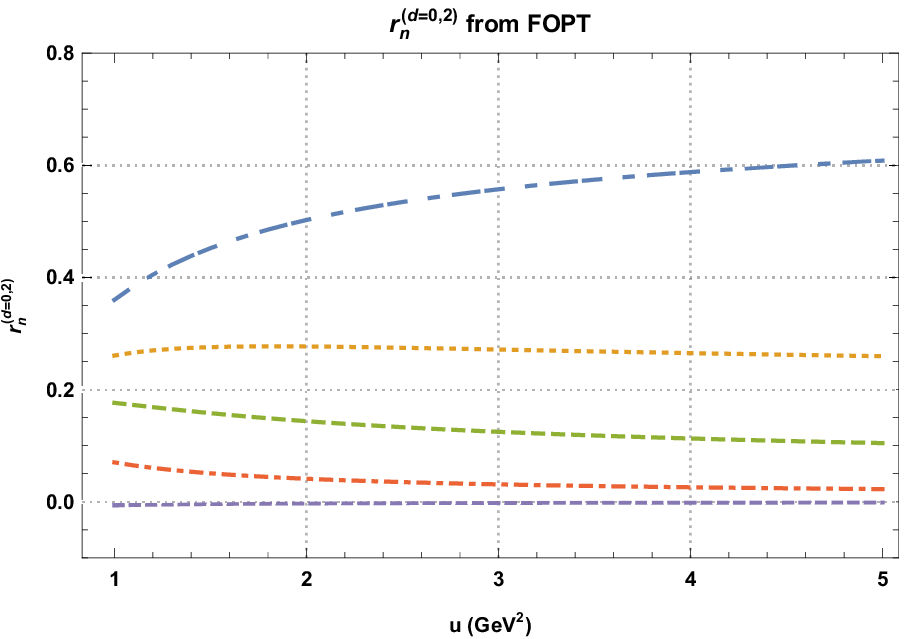}
		\includegraphics[width=.48\textwidth]{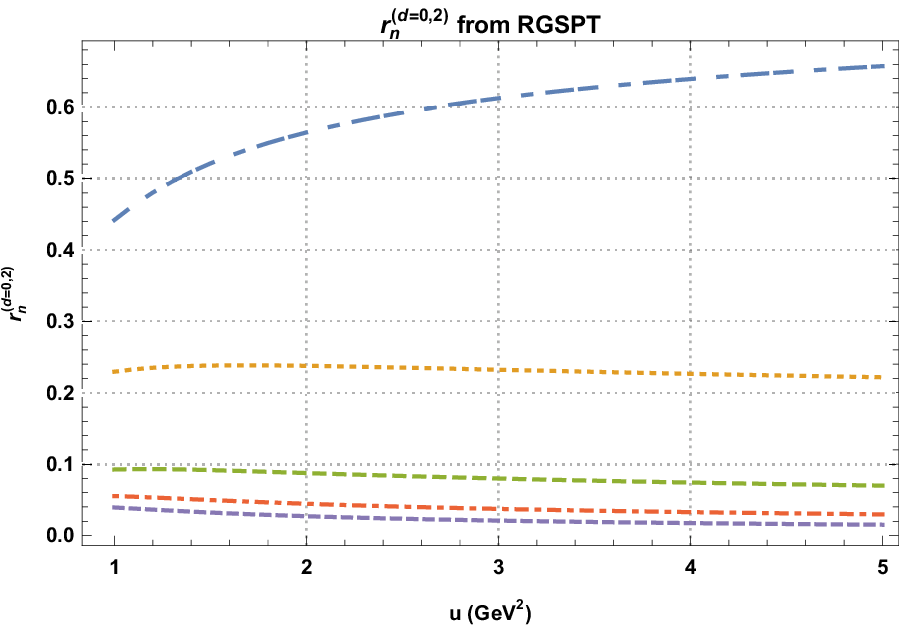}
 \caption{$r_n^{d=0,2}$ from FOPT and RGSPT. The lines from up to down correspond to $n=0,1,2,3,4$. }
 \label{fig:rn}
    \end{figure}

\subsection{Convergence of Borel transformed OPE using FOPT and RGSPT.}\label{subsec:conf_borel}
We use the ratio $r_n^{d=0,2}$, defined in Ref.~\cite{Chetyrkin:2005kn}, from Eq.~\eqref{eq:psi_Borel} as:
\begin{align}
    r_n^{d=0,2}(u)=\frac{\left(\frac{1}{u}\tilde{\Psi}''_0(u)+\frac{1}{u^2}\tilde{\Psi}''_2(u)\right)^{\left(\ordas{n}\right)}}{\Psi''(u)}\,.
\end{align}
The numerator in the above equation is evaluated using the contributions from $\ordas{n}$ from dimension-0 and dimension-2 corrections to $\Psi''(u)$. Using PDG values for the $m_s(2\GeV)=93.4\MeV$, $m_d(2\GeV)=4.67\MeV$ and setting $u=2.5\GeV^2$, we get the following contributions to $ r_{n}^{d=0,2}$:
\begin{align}
    r_{n}^{d=0,2}\vert_{\text{FOPT}}&=\{53.45\%, 27.46\%, 13.30\%, 3.51\%, -0.22\%\}\nonumber\,,\\
    r_{n}^{d=0,2}\vert_{\text{RGSPT}}&=\{59.25\%, 23.52\%, 8.34\%, 4.04\%, 2.35\%\}\,.\nonumber
\end{align}
From these numerical values, one may suspect that the FOPT has better convergence than RGSPT. This behavior can be attributed to the fact that there are large negative corrections from the Borel transform of the logarithmic  $\log^n\left(\mu^2/Q^2\right)$ terms as depicted in Fig.~\eqref{fig:A0}. The behavior of $r_n$ for different values of the Borel parameter can be found in Fig.~\eqref{fig:rn}.\par
These findings clearly show that RGSPT has the potential to reduce theoretical uncertainty significantly and has been the primary goal of this article.
\subsection{Instanton contribution} \label{subsec:insanton}
In addition to the OPE correction, the QCD vacuum structure becomes relevant at low energy, and contributions from the instantons become relevant at energy range$\sim \hs 1\GeV$. Their contributions are estimated using the instanton liquid model (ILM)~\cite{Ilgenfritz:1980vj,Shuryak:1981ff,Shuryak:1982dp} and are added to the pseudoscalar current correlator. These contributions are parameterized in terms of the instanton size $\rho_c$ and number density $n_c$. For the spectral density, we use the results from Ref.~\cite{Elias:1998fs,Maltman:2001gc,Yin:2021cbb}:
\begin{align}
    \rho^{\text{inst.}}_{i,j}&=\frac{1}{2\pi}\text{Im}(\Psi(s)_{\text{inst.}})\nonumber\\
    &=\frac{-3\eta_{ij}\left(m_i+m_j\right)^2}{4\pi}J_1\left(\rho_c \sqrt{s}\right) Y_1\left(\rho_c \sqrt{s}\right)\,,
    \label{eq:rho_insta}
\end{align}
where, $\rho_c=1/0.6$ and $\eta_{ud/us}=1/0.6$~\cite{Shuryak:1982dp}. In addition, we also need the Borel transform of the second derivative of the polarization function for instanton, which is given by~\cite{Narison:2014vka}:
\begin{align}
   \Psi_{5,ij}''(u)^{\text{inst.}}&=\hat{\mathcal{B}}_u[(\Psi_5''(s))_{\text{inst.}}]\nonumber\\&= \frac{3\eta_{ij} \rho_c^2\left(m_i+m_j\right)^2}{8 \pi ^2} e^{-\frac{1}{2} \rho_c^2\hs u} \nonumber\\&\bs\times\left[K_0\left(\frac{1}{2}\rho_c^2 u\right)+K_1\left(\frac{1}{2}\rho_c^2 u\right)\right]\,,
   \label{eq:borel_instal}
\end{align}
where $K_0$ and $K_1$ are the modified Bessel functions. These contributions are numerically relevant for low values of the Borel parameter $u\sim 1\GeV^2$.\par
Now, we have all the theoretical and phenomenological needed as input for the Borel sum rule in Eq.~\eqref{eq:bs_final}. In the next section, light quark mass determination using FOPT and RGSPT is performed. 
\section{Light quark mass determination}\label{sec:ms_md_determination}
In this section, we determine that masses of the strange quark mass using the Borel-Laplace sum rule in Eq.~\eqref{eq:bs_final} from the divergences of the axial vector current. It should be noted that the $m_u$ is determined using the ratio $\epsilon_{ud}\equiv m_u/m_d=0.474_{-0.074}^{+0.056}$~\cite{ParticleDataGroup:2022pth}.\par
Before moving to mass determination from the Sum rule, we need to fix the values for the continuum threshold $s_0$ and Borel parameter $u$. In principle, any determination from the Borel-Laplace sum rule should be independent of the choice of these parameters in the limit $u\gg s_0$. However, in practical cases, there is a dependence on the determinations of light quark masses on these parameters. For practical purposes, these parameters are tuned to get stable results for a given range. The Borel parameter is chosen large enough to suppress the contributions from non-perturbative condensate terms and resonances. However, the continuum threshold $s_0$ is chosen in a region where contributions from the higher resonances are negligible and spectral function can be approximated with the continuum pQCD correction. A proper window for $s_0$ and $u$ is crucial for the stable determination of the Borel-Laplace sum rule, and we have discussed them for FOPT and RGSPT and in this section for the individual as well as simultaneous $m_d$ and $m_s$ determination.\par
\begin{figure}[H]
	\centering
	\includegraphics[width=.48\textwidth]{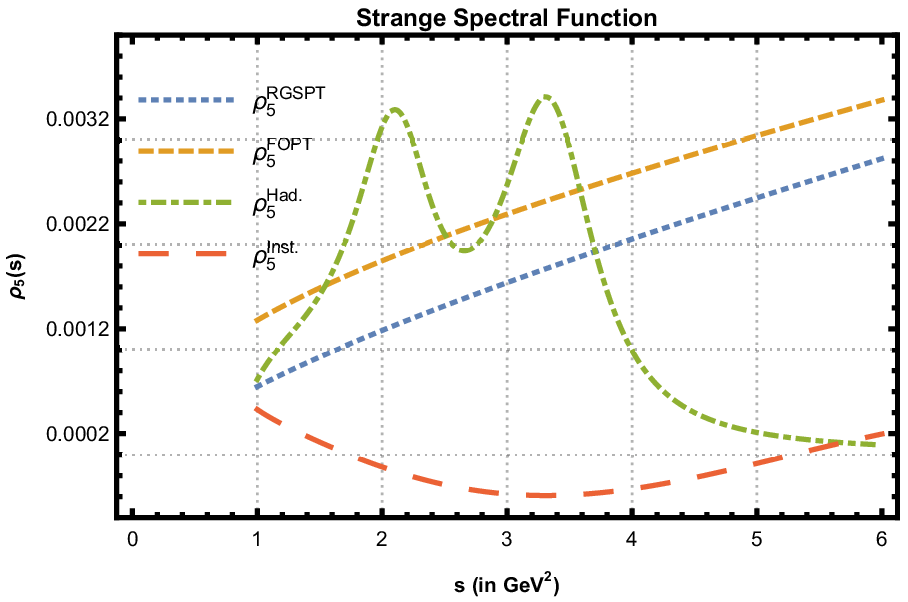}
	\includegraphics[width=.49\textwidth]{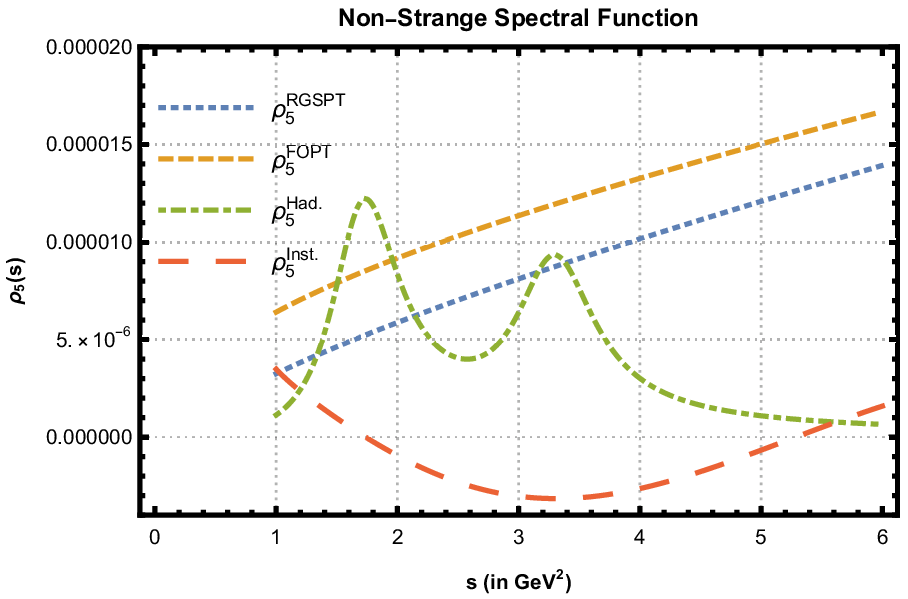}	
	\caption{Hadronic and theoretical spectral functions in the strange and non-strange channel.}
	\label{fig:rhoall}
\end{figure}

We can also perform quark mass determination by choosing the value of $s_0$ for which both hadronic and perturbative spectral functions are in agreement. However, these determinations are going to be very sensitive to the second resonance present in the hadronic spectral function. Another issue is the absence of information about higher resonances which are already neglected in this study. Various contributions to the spectral functions are presented in Figs.~\eqref{fig:rhoall} for non-strange and strange channels. For these channels, this agreement is found in the range $s_0\in \left[3.38,3.79\right]$ for which we have taken $s_0=3.58\pm 0.20\GeV^2$ in such determinations. However, such determinations are not taken in our final average due to the issues discussed above.

\subsection{\texorpdfstring{$m_s$}{} determination}\label{subsec:ms}
 To fix the free $s_0$ and $u$ parameters, we first perform $m_s$ determination at different values. In the FOPT prescription, there is a crossover around $s_0=4.5\hs\GeV^2$ for different values of the Borel parameter that can be seen in Fig.~\eqref{fig:msFOPTus01}. Therefore, we choose $s_0=4.5\pm0.5\GeV^2$ to minimize variation in the $m_s$ determination with respect to the Borel parameter from FOPT determinations. However, this is not the case for RGSPT as there is no crossing point in Fig.~\eqref{fig:msRGSPTus01}. There is a stability window for $s_0\in\hs\left[3.5,4.2\right]\GeV^2$ region for RGSPT, but we do not use this value as $\sqrt{s_0}$ is close to the mass of the second resonance. This results in slightly more uncertainty from the variation of $u$ in the $m_s$ determination compared to FOPT, which can be seen in Fig.~\eqref{fig:msmdu1}. However, we find that there is a linear increase in the difference of maximum and minimum values of strange quark mass ($\Delta(m_s)$) determination for $s_0\in\left[3,5\right]\GeV^2$ with $u\in\left[2,3\right]$, which can be seen in Fig.~\eqref{fig:delms}. This linear behavior is milder in the case of RGSPT compared to FOPT. \par
 Now, we move on to our final determination for which we adopt the choice of parameters used in Ref.~\cite{Chetyrkin:2005kn}. For the Borel parameter, we use $ u=2.5\pm0.5\GeV^2$ and the renormalization scale is varied in the range $u/2\leq\mu^2\leq2 u$. We take the continuum threshold value $s_0=4.5\pm0.5\hs\GeV^2$ and $m_u=2.16_{-0.26}^{+0.49}\MeV$~\cite{ParticleDataGroup:2022pth} as input in our determination. We obtain the following value of $m_s(2\GeV)$, using FOPT:
\begin{align}
  m_s(2\hs\GeV)=103.64_{-4.61}^{+6.45}\hs\MeV\,,
\end{align} 
and for RGSPT, we obtain:
\begin{align}
    m_s(2\hs\GeV)=104.20_{-4.29}^{+4.37}\hs\MeV\,.
\end{align}
The details of significant sources of uncertainties can be found in Table~\eqref{tab:md_ms_indiv_final}. The pQCD uncertainties containes uncertainties arising from uncertainties present in the quark and gluon condensates, $\as$, renormalization scale variation, and  tuncation uncertainty. The truncation uncertainty is calculated from the contribution of the last terms present in the expansion of $\as$ in the perturbative series.   Uncertainties from other parameters are included in the hadronic uncertainties.\par
 It is worth mentioning that the uncertainties coming from scale variation in RGSPT are significantly smaller than in FOPT, leading to small pQCD uncertainties compared to the hadronic uncertainties. It is important to note from Table.~\eqref{tab:md_ms_indiv_final} is that the total theoretical uncertainty from pQCD parameters is smaller than the hadronic uncertainties when RGSPT is used. We present the scale dependence in our determinations in Fig.~\eqref{fig:msscdep}. Another point to note is that the exclusion of the instanton term for the RGSPT and FOPT series leads to a decrease of strange quark mass about~$1.26\MeV$ and $1.24\MeV$, respectively. \par
Now, we also present our results for theoretical and hadronic spectral functions are in agreement. Using $s_0=3.58\pm 0.20\GeV^2$ and taking the rest of the parameters discussed above, we get the following determinations for FOPT and RGSPT scheme:
\begin{align}
m_s(2\GeV)&=107.29_{-5.83}^{+7.57}\MeV\,,\quad (\text{FOPT})\\
m_s(2\GeV)&=106.02_{-4.57}^{+4.36}\MeV\,,\quad (\text{RGSPT})
\end{align}
The dependence of these determinations on the Borel parameter is presented in Fig.~\eqref{fig:msmdQHu0}.
\subsection{\texorpdfstring{$m_d$}{} determination}\label{subsec:md}
Similar to $m_s$ determination, there is a crossover point for $m_d$ in FOPT, but near to $\pi(1800)$ resonance mass as we can see in Fig.~\eqref{fig:mds0u0}. Due to this, we choose $s_0=4.5\pm0.5\GeV^2$ as in the previous subsection. Using the same parameters and $\epsilon_{ud}$~\cite{ParticleDataGroup:2022pth} for FOPT, we obtain the following values:
\begin{align}
  m_d(2\hs\GeV)&=4.18_{-0.44}^{+0.51}\hs\MeV\,,\\
 \implies m_u(2\hs\GeV)&=1.98_{-0.37}^{+0.34}\hs\MeV\,,
\end{align}
and for RGSPT, we obtain the following value:
\begin{align}
    m_d(2\hs\GeV)&=4.21_{-0.39}^{+0.48}\hs\MeV\,.\\
    \implies m_u(2\hs\GeV)&=2.00_{-0.36}^{+0.33}\hs\MeV\,.
\end{align}
 We present the scale dependence in our determinations in Fig.~\eqref{fig:mdscdep}. The details of the sources of uncertainties can be found in Table~\eqref{tab:md_ms_indiv_final}. Exclusion of the instanton terms leads to a decrease in the central value of $m_d(2\GeV)$ by~$0.20\MeV$ and~$0.13\MeV$ in determinations using FOPT and RGSPT prescriptions, respectively.\par
\begin{figure}[H]
	\centering
	\centering
	\includegraphics[width=\linewidth]{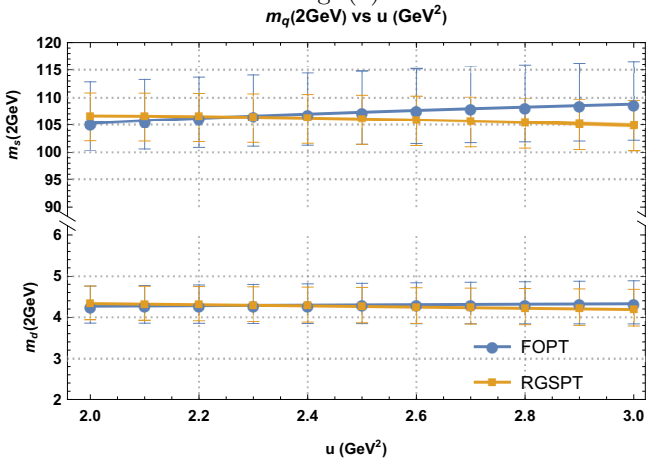}
	\caption{\label{fig:msmdQHu0} Borel parameter dependence of the individual determinations of $m_d(2\GeV)$ assuming quark hadron duality is obeyed at $s_0=$ from FOPT and RGSPT prescriptions.}
\end{figure}

Now, using $s_0=3.58\pm 0.20\GeV^2$ and taking the rest of the parameters discussed above, we get the following determinations for FOPT and RGSPT scheme:
	\begin{align}
		m_d(2\GeV)&=4.30_{-0.46}^{+0.52}\MeV\,,\quad (\text{FOPT})\\
		\implies m_u(2\GeV)&=2.04_{-0.40}^{+0.35}\MeV\,,\\
		m_d(2\GeV)&=4.26_{-0.39}^{+0.46}\MeV\,,\quad (\text{RGSPT})\\
		\implies m_u(2\GeV)&=2.02_{-0.40}^{+0.32}\MeV\,,\,.
	\end{align}
The dependence of these determinations on the Borel parameter can be found in Fig.~\eqref{fig:msmdQHu0}.

\subsection{Simultaneous \texorpdfstring{$m_d$ and $m_s$}{} determination.}
We can also perform the simultaneous determination of the $m_s$ and $m_d$ using the sum rule in Eq.~\eqref{eq:bs_final} for strange and non-strange channels. Using FOPT, we obtain:
\begin{align}  
    m_s(2\hs\GeV)&=103.80_{-4.22}^{+6.14}\hs\MeV\,,\\
    m_d(2\hs\GeV)&=4.18_{-0.44}^{+0.50}\hs\MeV\,,\\
     \implies m_u(2\hs\GeV)&=1.98_{-0.37}^{+0.33}\hs\MeV\,,
\end{align} 
and for RGSPT, we obtain the following values:
\begin{align}
        m_s(2\hs\GeV)&=104.34_{-4.24}^{+4.32}\hs\MeV\,,\\
        m_d(2\hs\GeV)&=4.21_{-0.45}^{+0.48}\hs\MeV\,,\\
        \implies m_u(2\hs\GeV)&=2.00_{-0.38}^{+0.33}\hs\MeV\,.
\end{align}
The details of sources of uncertainties in the determination of quark masses can be found in Table~\eqref{tab:mdms_final}. In this case, uncertainty in $m_s$ determination is smaller than one obtained in subsection~\eqref{subsec:ms}. The values obtained for $m_s$ and $m_d$ are very close to the determination from subsection~\eqref{subsec:ms} and ~\eqref{subsec:md}. \par
Using $s_0=3.58\pm 0.20\GeV^2$ for FOPT, we obtain:
	\begin{align}  
		m_s(2\hs\GeV)&=107.39_{-5.08}^{+6.95}\hs\MeV\,,\\
		m_d(2\hs\GeV)&=4.30_{-0.44}^{+0.51}\hs\MeV\,,\\
		\implies m_u(2\hs\GeV)&=2.04_{-0.40}^{+0.34}\hs\MeV\,,
	\end{align} 
	and for RGSPT, we obtain the following values:
	\begin{align}
		m_s(2\hs\GeV)&=106.14_{-4.52}^{+4.32}\hs\MeV\,,\\
		m_d(2\hs\GeV)&=4.26_{-0.43}^{+0.46}\hs\MeV\,,\\
		\implies m_u(2\hs\GeV)&=2.02_{-0.40}^{+0.32}\hs\MeV\,.
\end{align}

\begin{widetext}

\begin{table}[H]
    \centering
    \begin{tabular}{||c|c|c|c|c|c|c|c|c|c|c|c|c|c|c||}\hline\hline
    \text{} &\multicolumn{7}{c|}{\textbf{FOPT}}&\multicolumn{7}{c|}{\textbf{RGSPT}}\\\cline{2-15}
\text{Quark mass}&Final Value&$\mu$&$\as$&u&$s_0$&\text{pQCD}&\text{Had.}&Final Value&$\mu$&$\as$&u&$s_0$&\text{pQCD}&\text{Had.}
\\\hline\hline
\multirow{2}{5em}{$m_d(2\GeV)$}&\multirow{2}{5em}{$4.18^{+0.51}_{-0.44}$}&+0.18&+0.06&+0.07&+0.08&+0.20&+0.47&\multirow{2}{5em}{$4.21^{+0.48}_{-0.39}$}&+0.02&+0.06&+0.10&+0.05&+0.10&+0.47\\\cline{3-8}\cline{10-15}
\text{}&\text{}&-0.07&-0.06&-0.05&-0.09&-0.11&-0.43&\text{}&-0.01&-0.06&-0.08&-0.07&-0.10&-0.38\\\hline\hline\hline
\multirow{2}{5em}{$m_s(2\GeV)$}&\multirow{2}{5em}{$103.64^{+6.45}_{-4.61}$}&+4.59&+1.40&+0.76&+2.59&+5.14&+3.91&\multirow{2}{5em}{$104.20^{+4.37}_{-4.29}$}&+0.55&+1.42&+1.40&+1.76&+2.45&+3.61\\\cline{3-8}\cline{10-15}
\text{}&\text{}&-1.66&-1.40&-0.52&-2.66&-2.85&-3.62&\text{}&-0.52&-1.40&-1.48&-2.13&-2.43&-3.54\\\hline
    \end{tabular}
    \caption{\label{tab:md_ms_indiv_final} $m_d$ and $m_s$ determination using FOPT and RGSPT and the sources of uncertainties denoted in the column.}
\end{table}
\begin{table}[ht]
    \centering
    \begin{tabular}{||c|c|c|c|c|c|c|c|c|c|c|c|c|c|c||}\hline\hline
    \text{} &\multicolumn{7}{c|}{\textbf{FOPT}}&\multicolumn{7}{c|}{\textbf{RGSPT}}\\\cline{2-15}
\text{Quark mass}&Final Value&$\mu$&$\as$&u&$s_0$&\text{pQCD}&\text{Had.}&Final Value&$\mu$&$\as$&u&$s_0$&\text{pQCD}&\text{Had.}
\\\hline\hline
\multirow{2}{5em}{$m_d(2\GeV)$}&\multirow{2}{5em}{$4.18^{+0.50}_{-0.44}$}&+0.18&+0.06&+0.07&+0.08&+0.19&+0.47&\multirow{2}{5em}{$4.21^{+0.48}_{-0.45}$}&+0.02&+0.06&+0.10&+0.05&+0.10&+0.47\\\cline{3-8}\cline{10-15}
\text{}&\text{}&-0.07&-0.06&-0.05&-0.09&-0.09&-0.43&\text{}&-0.01&-0.06&-0.08&-0.07&-0.10&-0.43\\\hline\hline
\multirow{2}{5em}{$m_s(2\GeV)$}&\multirow{2}{5em}{$103.80^{+6.14}_{-4.22}$}&+4.54&+1.38&+0.72&+2.56&+4.77&+3.87&\multirow{2}{5em}{$104.34^{+4.32}_{-4.24}$}&+0.55&+1.41&+1.34&+1.74&+2.42&+3.57\\\cline{3-8}\cline{10-15}
\text{}&\text{}&-1.66&-1.38&-0.49&-2.62&-2.24&-3.58&\text{}&-0.52&-1.38&-1.44&-2.11&-2.40&-3.50\\\hline
    \end{tabular}
    \caption{\label{tab:mdms_final} $m_d$ and $m_s$ determination using FOPT and RGSPT and the sources of uncertainties denoted in the column.}
\end{table}

\begin{figure}[ht]
\centering
	\begin{subfigure}{.48\textwidth}
		\centering
		\includegraphics[width=\linewidth]{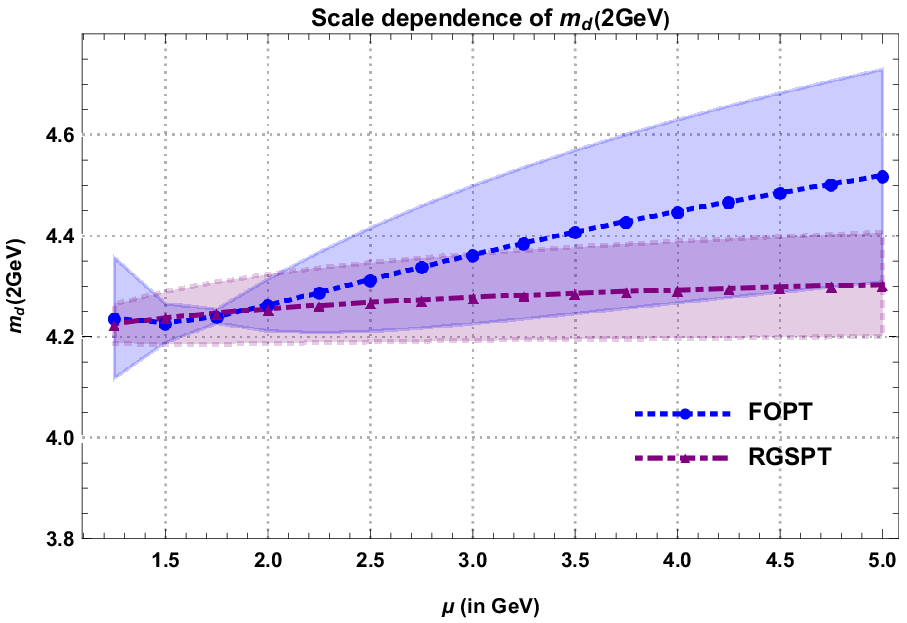}
 \caption{ \label{fig:mdscdep}}
	\end{subfigure}
	\begin{subfigure}{.48\textwidth}
		\centering
		\includegraphics[width=\linewidth]{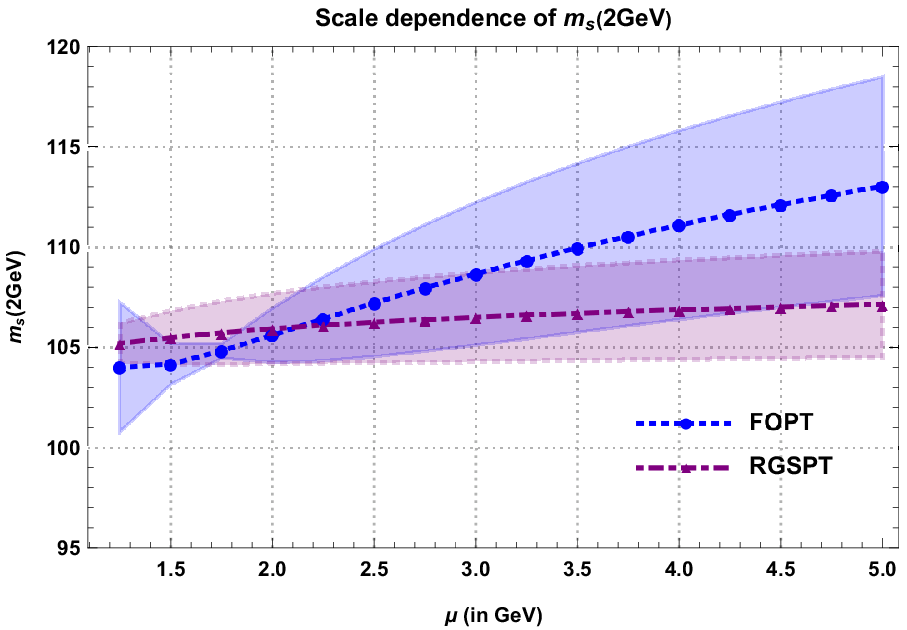}
\caption{  \label{fig:msscdep}}
 \end{subfigure}

    \caption{The scale dependence in the individual $m_d$ and $m_s$ determinations using FOPT and RGSPT. The bands in the plot represent truncation uncertainty at different scales.}
    \label{fig:msmdscdep}
\end{figure}

\begin{figure}[ht]
\centering
	\begin{subfigure}{.48\textwidth}
		\centering
		\includegraphics[width=\linewidth]{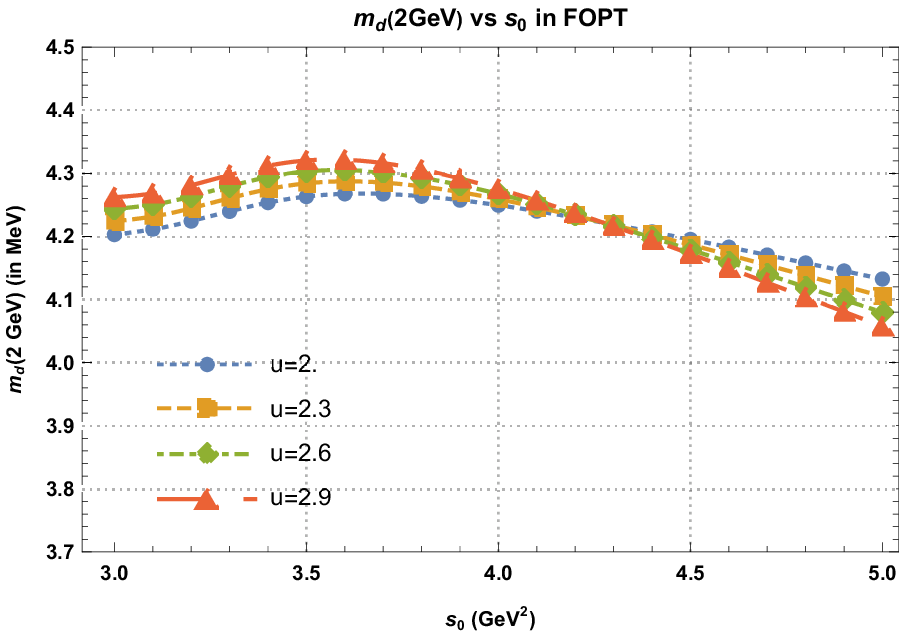}
 \caption{ \label{fig:msFOPT1}}
	\end{subfigure}
	\begin{subfigure}{.48\textwidth}
		\centering
		\includegraphics[width=\linewidth]{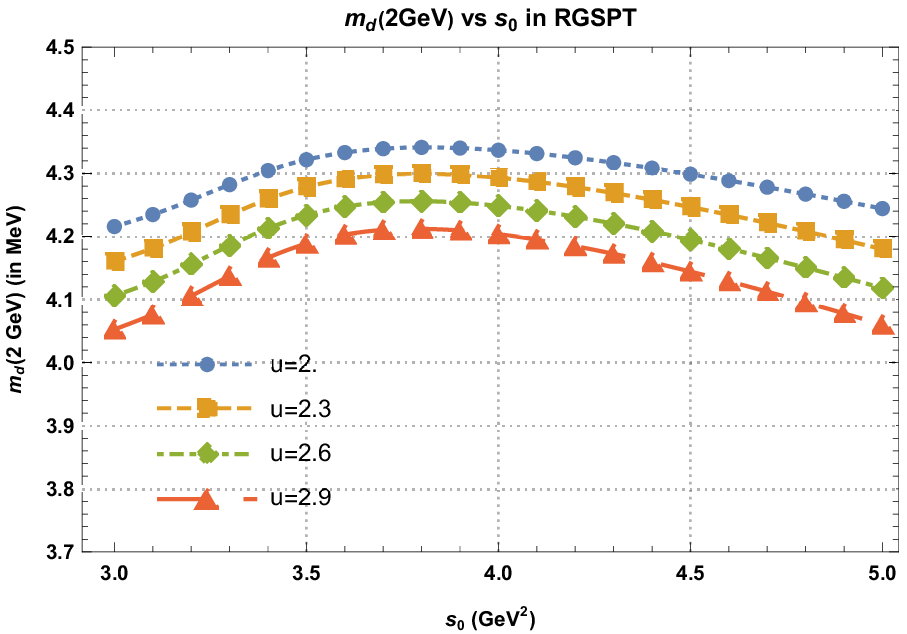}
\caption{  \label{fig:mdRGSPT1}}
 \end{subfigure}
    \caption{$m_d(2\GeV)$ calculated at different values of Borel parameter and $s_0$ using: \eqref{fig:msFOPT1} FOPT and, \eqref{fig:mdRGSPT1} RGSPT prescription.}
    \label{fig:mds0u0}
\end{figure}

\begin{figure}[H]
\centering
	\begin{subfigure}{.48\textwidth}
		\centering
		\includegraphics[width=\linewidth]{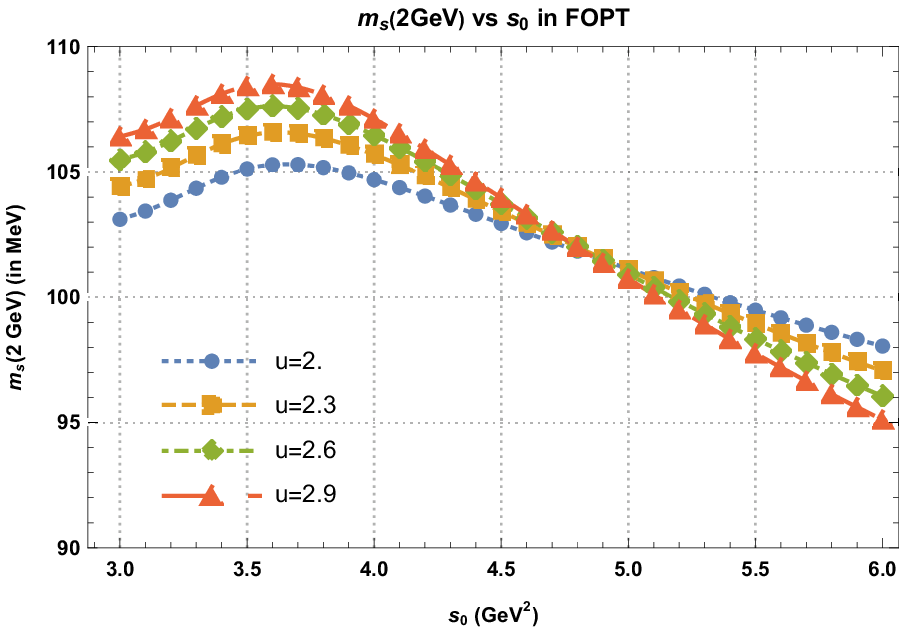}
 \caption{ \label{fig:msFOPTus01}}
	\end{subfigure}
	\begin{subfigure}{.48\textwidth}
		\centering
		\includegraphics[width=\linewidth]{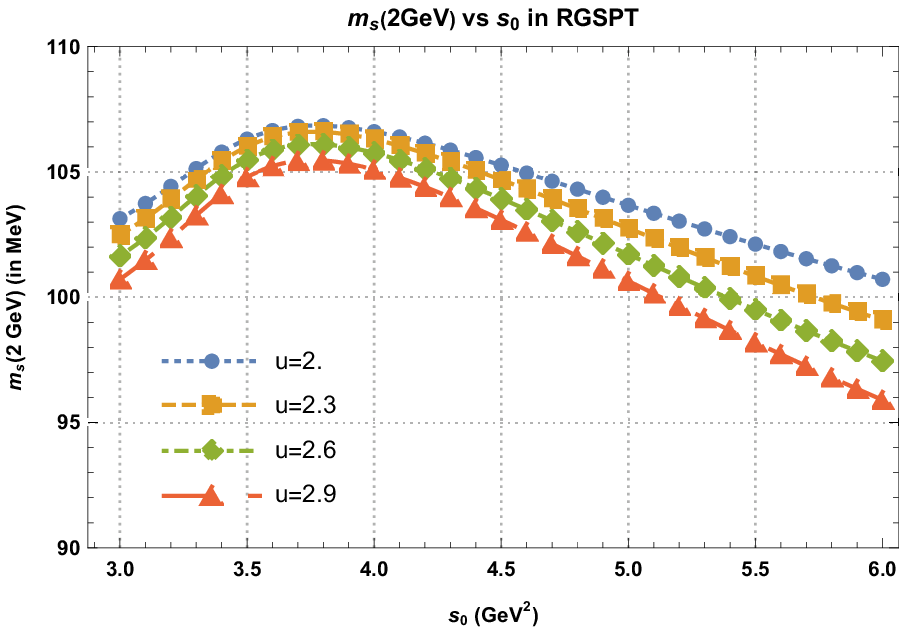}
\caption{  \label{fig:msRGSPTus01}}
 \end{subfigure}
 \newline
  \begin{subfigure}{.48\textwidth}
		\centering
		\includegraphics[width=\linewidth]{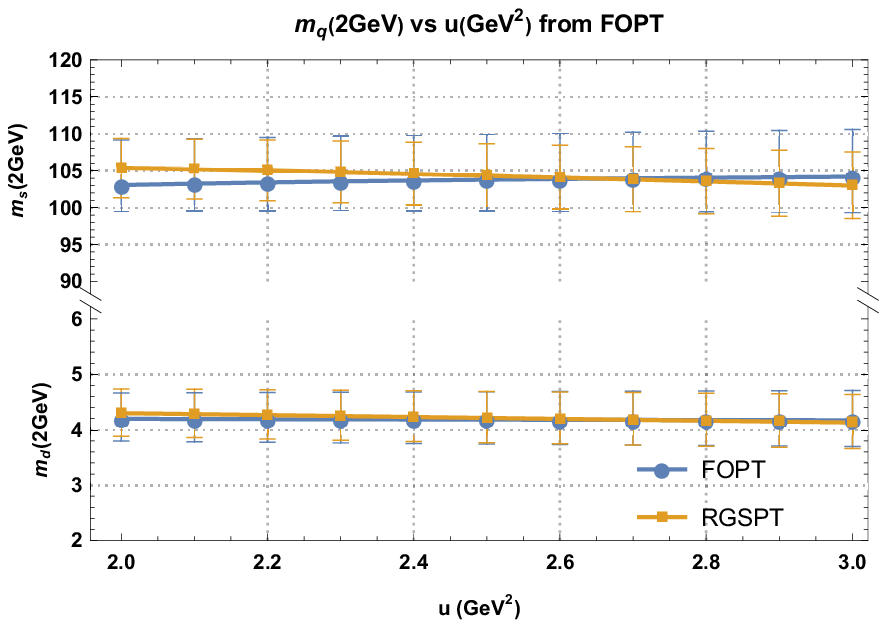}
\caption{ \label{fig:msmdu1}}
 \end{subfigure}
 \begin{subfigure}{.48\textwidth}
		\centering
		\includegraphics[width=\linewidth]{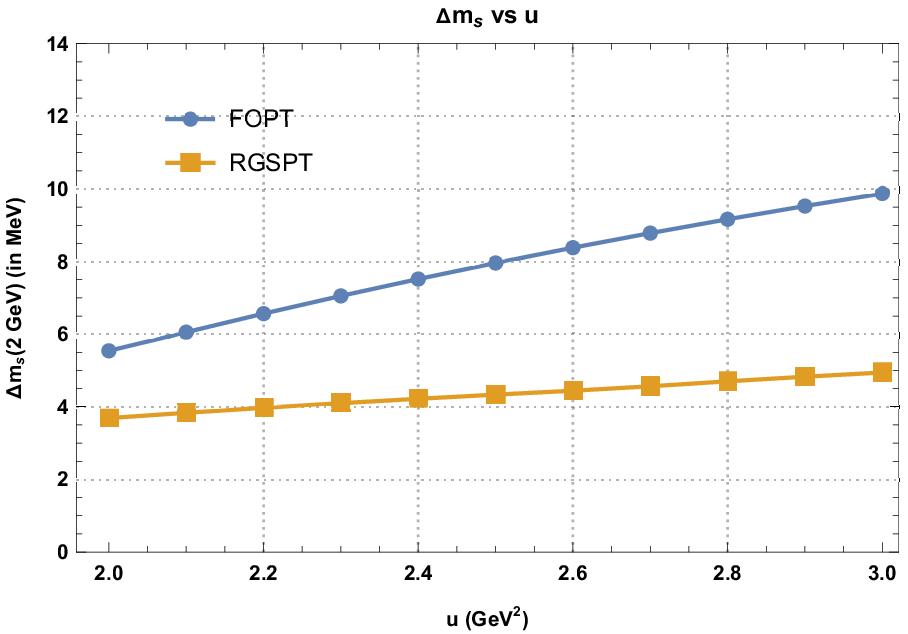}
\caption{  \label{fig:delms}}
 \end{subfigure}
    \caption{$m_s(2\GeV)$ calculated at different values of Borel parameter and $s_0$ using: \eqref{fig:msFOPTus01} FOPT and, \eqref{fig:msRGSPTus01} RGSPT scheme. In Fig.~\eqref{fig:msmdu1}, $m_s(2\GeV)$ and $m_d(2\GeV)$ at different values of the Borel parameter in the range $u\in\left[2,3\right]\GeV^2$. In \eqref{fig:delms}, $\Delta m_s(2\GeV)$ obtained by varying $s_0\in\left[3,5\right]\GeV^2$ at different values of Borel parameter $u$.}
    \label{fig:msFOPTus0}
\end{figure}

\end{widetext}

\section{Summary and conclusion}\label{sec:summary}
We have used the Borel-Laplace sum rule to determine the light quark masses from the correlator of the divergence of the axial vector current. The sum rule uses both hadronic as well as perturbative contributions. In section~\eqref{sec:rgspt}, we briefly review the procedure of RG summation in RGSPT and its importance in the RG improvement for the theoretical quantities used in the Borel-Laplace sum rule. \par In section~\eqref{sec:hadronic_info}, we discuss the hadronic pseudoscalar spectral function for which no experimental information is available. However, these contributions can be parameterized in terms of the information available on the masses and decay width of the spectral function. We use commonly used hadronic parametrization from Dominguez and Rafael~\cite{Dominguez:1986aa} in this article for the light quark mass determination, and it has a good agreement with another parametrization by Maltman and Kambor~\cite{Maltman:2001gc}, which can be seen in Fig.~\eqref{fig:had}.\par
In section\eqref{sec:OPE}, the continuum contributions are discussed in detail. The most commonly used FOPT prescription results are already available in the literature. These determinations have large uncertainties from the variation of the renormalization scale. RGSPT can reduce such uncertainties and is inspired by the findings of Ref.~\cite{AlamKhan:2023dms}, we first sum the kinematical $\pi^2$-terms appearing due to analytic continuation of the spectral function in section~\eqref{subsec:ancont}. The analytic continuation using RGSPT also results in better convergence of the perturbation series for the dimension-0 contribution. We also find better convergence and improved scale dependence for the $\Psi''(Q^2)$, which is also an RG invariant quantity. These improvements can be seen in Fig.~\eqref{fig:scdepR0Pd2}. \par 
In section~\eqref{subsec:BorelTransform}, we first calculate the Borel transformation for the $\Psi''(Q^2)$ in the RGSPT prescription. The Borel transformation for RGSPT can only be performed numerically, and it resums all the Euler's constant and various Zeta functions that arise due to Borel transformation for RGSPT. The FOPT results are found to be poorly convergent and oscillate around the all-order result from the RGSPT. RGSPT also improves the scale dependence of the Borel transformed $\Psi''(Q^2)$ which is used as input in the Borel-Laplace sum rule in Eq.~\eqref{eq:bs_final}. These improvements can be seen in Figs.~\eqref{fig:A0},\eqref{fig:RGP2u}. The result obtained is used in section~\eqref{subsec:conf_borel} to test the convergence of $\Psi''(u)$. FOPT series is found to be slightly more convergent than RGSPT, but it is argued that RGSPT results are more trustworthy as Borel transformation in FOPT has oscillatory behavior for the known results. We also include small instanton contributions using results from ILM in section~\eqref{subsec:insanton}.\par
We determine the light quark masses in section~\eqref{sec:ms_md_determination} using the free parameters $s_0$ and $u$ used in Ref.~\cite{Chetyrkin:2005kn}. This particular choice leads to small $u$ dependence in the FOPT determinations of the $m_s$, which can be seen in Fig.~\eqref{fig:msFOPTus01}. For the RGSPT determination, the stability region is closer to the second resonance; therefore, we have used the choices for these free parameters from FOPT. This leads to slightly large uncertainty in the $m_s$ determination from the variations of $u$. In addition to the individual determination of the $m_s$ and $m_d$, we have performed simultaneous determination and found a slightly more precise value for $m_s$. These results are presented in Table~\eqref{tab:md_ms_indiv_final} and Table~\eqref{tab:mdms_final}. In addition to this, we have also presented our determination by choosing $s_0=3.58\pm0.20\GeV^2$ in the resonance region where there is good agreement between theoretical and hadronic spectral function. These values result in higher values of the quark masses which can be seen in Fig.~\eqref{fig:msFOPTus0} and Fig.~\eqref{fig:msFOPTus0}. Since this value choice of $s_0$ is sensitive to the parameters of the second resonance and higher resonances are neglected, we do not consider them in our final determinations. \par
Now, we give our final determination for the light quark masses, which comes from the simultaneous determination of the $m_s$ and $m_d$ and their values at $2\GeV$ are:
\begin{align}
        m_s(2\hs\GeV)&=104.34_{-4.24}^{+4.32}\hs\MeV\,,\\
        m_d(2\hs\GeV)&=4.21_{-0.45}^{+0.48}\hs\MeV\,,\\
        m_u(2\hs\GeV)&=2.00_{-0.40}^{+0.33}\hs\MeV\,.
\end{align}
 Corresponding PDG average values~\cite{ParticleDataGroup:2022pth} are:
\begin{align}
        m_s(2\hs\GeV)&=93.4_{-3.4}^{+8.6}\hs\MeV\,,\\
        m_d(2\hs\GeV)&=4.67_{-0.17}^{+0.48}\hs\MeV\,,\\
        m_u(2\hs\GeV)&=2.16_{-0.26}^{+0.49}\hs\MeV\,.
\end{align}
which is in agreement with our findings. 
\vspace{1cm}
\section*{Acknowledgment}
We thank Mr. Aadarsh Singh for carefully reading the manuscript and for his valuable comments. We are also very thankful to Prof. Apoorva Patel for the financial support. We thank anonymous referee for valuable comments and suggestions that has greatly improved this article. The author is also supported by a scholarship from the Ministry of Human Resource Development (MHRD), Govt. of India. This work is a part of the author's Ph.D. thesis.

	\appendix
 
	\begin{widetext}
		
		\section{Running of the Strong Coupling and the Quark Masses in the pQCD. \label{app:mass_run}}
		The running of strong coupling and the quark masses are computed by solving the following differential equations:
		\begin{align}
			\mu^2\frac{d}{d\mu^2}x(\mu)=\beta(x)=-\sum_i x^{i+2} \beta_i\,, \quad
			\mu^2 \frac{d}{d\mu^2}m(\mu)\equiv&\hspace{2mm}m \hspace{.4mm}\gamma_m =	-m\sum_{i}\gamma_i \hspace{.4mm}x^{i+1} \label{anomalous_dim}\,.
		\end{align}
		where $\beta_i$ are the QCD beta function coefficients and $\gamma_i$ are the quark mass anomalous dimension.\par 
		The QCD beta function coefficients are known to five-loop \cite{vanRitbergen:1997va,Gross:1973id,Caswell:1974gg, Jones:1974mm,Tarasov:1980au,Larin:1993tp,Czakon:2004bu,Baikov:2016tgj,Herzog:2017ohr}  and their analytic expression for $n_f$-active flavour are: 
		\begin{align}
			&\beta_0 = \frac{11}{4}-\frac{1}{6}n_f\, \quad 	\beta_1= \frac{51}{8} - \frac{19}{24}n_f\,,\quad
			\beta_2 = \frac{2857}{128} - \frac{5033}{1152} n_f + \frac{325}{3456}n_f^2\,, \nonumber\\
			&\beta_3 = \frac{149753}{1536} - \frac{1078361}{41472} n_f + \frac{50065}{41472} n_f^2 + \frac{1093}{186624} n_f^3 + \frac{891}{64} \zeta(3) - \frac{1627}{1728} n_f \zeta(3) + \frac{809}{2592} n_f^2 \zeta(3)\nonumber\,,
		\end{align}
		\begin{align}
			\beta_{4} =&\frac{8157455}{16384}+\frac{621885 \zeta (3)}{2048}-\frac{9801 \pi ^4}{20480} -\frac{144045 \zeta (5)}{512}+ n_{f}\big(-\frac{336460813}{1990656}-\frac{1202791 \zeta (3)}{20736}+ \frac{6787 \pi ^4}{110592}+\frac{1358995 \zeta (5)}{27648} \big)& \nn \\ &+n_{f}^{2}\left(\frac{25960913}{1990656}+\frac{698531 \zeta (3)}{82944}-\frac{5263 \pi ^4}{414720}-\frac{5965 \zeta (5)}{1296}\right)+n_{f}^{3}\left(-\frac{630559}{5971968}-\frac{24361 \zeta (3)}{124416}+\frac{809 \pi ^4}{1244160}+\frac{115 \zeta (5)}{2304}\right)& \nn \\ &+n_{f}^{4}\left(\frac{1205}{2985984}-\frac{19 \zeta (3)}{10368}\right)\, .
		\end{align}			
		The known five-loop quark mass anomalous dimension coefficients\cite{Tarrach:1980up,Tarasov:1982plg,Larin:1993tq,Vermaseren:1997fq,Chetyrkin:1997dh,Baikov:2014qja,Luthe:2016ima,Luthe:2016xec} are:
		\begin{align}
			\gamma_m^{(0)}=&1 \,, \quad \gamma_m^{(1)}=\frac{1}{4^2} \left(\frac{202}{3} +\frac{-20}{9} n_f\right)\,, \quad				\gamma_m^{(2)}=\frac{1}{4^3}\left(1249+n_f \left(-\frac{160 \zeta (3)}{3}-\frac{2216}{27}\right)-\frac{140 n_f^2}{81}\right)\nonumber\\
			\gamma_m^{(3)}=&\frac{1}{4^4} \Bigg(\frac{135680 \zeta (3)}{27}-8800 \zeta (5)+\frac{4603055}{162}+n_f \left(-\frac{34192 \zeta (3)}{9}+880 \zeta (4)-\frac{18400 \zeta (5)}{9}-\frac{91723}{27}\right)\nonumber\\&+n_f^2 \left(\frac{800 \zeta (3)}{9}-\frac{160 \zeta (4)}{3}+\frac{5242}{243}\right)+n_f^3 \left(\frac{64 \zeta (3)}{27}-\frac{332}{243}\right)\Bigg)\nonumber\\
			\gamma_m^{(4)} = & \frac{1}{4^5}\Bigg(\frac{99512327}{162}+\frac{46402466 \zeta (3)}{243}+96800 \zeta (3)^2-\frac{698126 \zeta (4)}{9}-\frac{231757160 \zeta (5)}{243}+242000 \zeta (6)+412720 \zeta (7)\nonumber \\ &+n_f \Big(-\frac{150736283}{1458}-\frac{12538016 \zeta (3)}{81}-\frac{75680 \zeta (3)^2}{9}+\frac{2038742 \zeta (4)}{27}+\frac{49876180 \zeta (5)}{243}-\frac{638000 \zeta (6)}{9}\nonumber\\&-\frac{1820000 \zeta (7)}{27}\Big)+n_f^2 \left(+\frac{1320742}{729}+\frac{2010824 \zeta (3)}{243}+\frac{46400 \zeta (3)^2}{27}-\frac{166300 \zeta (4)}{27}-\frac{264040 \zeta (5)}{81}+\frac{92000 \zeta (6)}{27}\right)\nonumber\\&+n_f^3 \left(\frac{91865}{1458}+\frac{12848 \zeta (3)}{81}+\frac{448 \zeta (4)}{9}-\frac{5120 \zeta (5)}{27}\right)+n_f^4 \left(-\frac{260}{243}-\frac{320 \zeta (3)}{243}+\frac{64 \zeta (4)}{27}\right)\Bigg)\,.
		\end{align}
  \section{Contribution to current correlator } \label{app:dim0adler}
		\subsection{Dimension zero contributions}
  The zero-dimensional contribution to OPE is known to five loops ($\as^4$)~\cite{Gorishnii:1990zu,Chetyrkin:1996sr,Baikov:2005rw}. We are using the following expression for $\Psi_0$:
\begin{align}
    \Psi_0(q^2)&=\frac{3 }{8 \pi ^2}\Bigg\lbrace L+\left(L^2+\frac{17 L}{3}\right) x+x^2 \left(\frac{17 L^3}{12}+\frac{95 L^2}{6}+L \left(\frac{9631}{144}-\frac{35 \zeta (3)}{2}\right)\right)\nonumber\\&+x^3 \Big[L^2 \left(\frac{4781}{18}-\frac{475 \zeta (3)}{8}\right)+\frac{221 L^4}{96}+\frac{229 L^3}{6}+L \left(-\frac{91519 \zeta (3)}{216}+\frac{715 \zeta (5)}{12}-\frac{\pi ^4}{36}+\frac{4748953}{5184}\right)\Big]\nonumber\\&+x^4 \Big[L \left(\frac{192155 \zeta (3)^2}{216}-\frac{46217501 \zeta (3)}{5184}+\frac{455725 \zeta (5)}{432}-\frac{52255 \zeta (7)}{256}-\frac{125 \pi ^6}{9072}-\frac{3491 \pi ^4}{10368}+\frac{7055935615}{497664}\right)\nonumber\\&+L^2 \left(-\frac{1166815 \zeta (3)}{576}+\frac{24025 \zeta (5)}{96}-\frac{\pi ^4}{36}+\frac{97804997}{20736}\right)+L^3 \left(\frac{3008729}{3456}-\frac{5595 \zeta (3)}{32}\right)+\frac{51269 L^4}{576}+\frac{1547 L^5}{384}\Big]\Bigg\rbrace\,,
\end{align}
where $x\equiv \as(\mu))/\pi $ and $L=\log(\frac{\mu^2}{-q^2})$. This expression reproduces the results for $\mathcal{R}_0$ and $\tilde{\Psi}_0''(u)$ in Ref.~\cite{Chetyrkin:2005kn}.\par 
For RGSPT, these quantities can be derived using Eq.~\eqref{eq:master_relation} and Eq.~\eqref{eq:def_Adler} from the Adler function. The RGSPT expression for the dimension-0 Adler function is given by:
		\begin{align}
			\mathcal{D}_0(q^2)=&\frac{3 \left(m_{i}+m_{j}\right)^2 }{8 \pi ^2 w^{\frac{8}{9}}}\Bigg\lbrace1+x \left(-1.790+\frac{7.457}{w}-\frac{1.580 \log (w)}{w}\right)+x^2 \Big(-0.339+\frac{60.699}{w^2}+\frac{2.653 \log ^2(w)}{w^2}-\frac{14.514}{w}\nonumber\\&+\left(\frac{2.829}{w}-\frac{27.849}{w^2}\right) \log (w)\Big)+x^3 \Big(-\frac{129.189}{w^2}+\frac{599.649}{w^3}-\frac{4.542 \log ^3(w)}{w^3}+\left(\frac{76.231}{w^3}-\frac{4.750}{w^2}\right) \log ^2(w)\nonumber\\&+\left(\frac{53.766}{w^2}-\frac{361.248}{w^3}+\frac{0.536}{w}\right) \log (w)-\frac{5.207}{w}+0.593\Big)+x^4 \Big(-12.673+\frac{15.012}{w}-\frac{66.312}{w^2}-\frac{1339.755}{w^3}\nonumber\\&+\frac{6992.440}{w^4}+\frac{7.851 \log ^4(w)}{w^4}+\left(\frac{8.131}{w^3}-\frac{183.752}{w^4}\right) \log ^3(w)+\left(-\frac{146.509}{w^3}+\frac{1384.280}{w^4}-\frac{0.901}{w^2}\right) \log ^2(w)\nonumber\\&+\left(\frac{18.438}{w^2}+\frac{759.073}{w^3}-\frac{4787.937}{w^4}-\frac{0.938}{w}\right) \log (w)\Big)\Bigg\rbrace\,,
		\end{align}
	where, $w=1-x(\mu)\beta_0 \log(\frac{\mu^2}{-q^2})$. 

		\subsection{Dimension-2 Corrections\label{app:dim2adler}}
  The dimension-2 contributions with full mass dependence are available to $\ordas{1}$ in Refs.~\cite{Chetyrkin:1985kn,Generalis:1990id,Jamin:1992se,Jamin:1994vr}. Additional $\ordas{2}$ correction is taken from Ref.~\cite{Chetyrkin:2005kn}. For FOPT, we use the following expression for dimension-2 contribution to the current correlator: 
	\begin{align}
	    \Psi_2=&\frac{3 }{8 \pi ^2} \Bigg\lbrace\left(m_{i}^2+m_{j}^2\right) \left(x^2 \left(-\frac{25 L^3}{3}-\frac{97 L^2}{2}+L \left(\frac{154 \zeta (3)}{3}-\frac{5065}{36}\right)\right)+\left(-4 L^2-\frac{32 L}{3}\right) x-2 L\right)\nonumber\\&-m_{i} m_{j} \left(x \left(-4 L^2-\frac{56 L}{3}+8 \zeta (3)-\frac{88}{3}\right)-2 L-4\right)\Bigg\rbrace\,.
	\end{align}
 For RGSPT, we derive the spectral function from the dimension-2 Adler function:
 \begin{align}
     \mathcal{D}_2=&\frac{\left(m_{i}+m_{j}\right)^2\left(m_{i}^2+m_{j}^2\right)  }{139968 \pi ^2 w^{34/9}}\Bigg(8 x \left(729 \left(20 w^2-251\right) x \zeta (3)+32 \log (w) (8 w (290 x-81)+1600 x \log (w)-12607 x)\right)\nonumber\\&\bs\bs+52488 w^2+(5497360-w (7643 w+1797332)) x^2+1296 (361-145 w) w x\Bigg)\nonumber\\&+\frac{\left(m_{i}+m_{j}\right)^2 m_{i} m_{j} }{216 \pi ^2 w^{25/9}} (w (290 x-81)+256 x \log (w)-1046 x)
 \end{align}

\subsection{Dimension-4 contributions}
\label{app:dim4}
The dimension-4 contributions can be obtained from Ref.~\cite{Pascual:1981jr,Jamin:1992se,Jamin:1994vr}. For dimension-4 contributions, we use an RG-invariant combination of the condensates given in Ref.~\cite{Spiridonov:1988md,Baikov:2018nzi}. The constant terms at this order are important for the Borel-Laplace operator. We give a summed expression for the current correlator:
\begin{align}
\Psi_4=&-\frac{1}{162 w^{17/9}} (w (145 x-81)+128 x \log (w)-442 x) \langle \sum_{i=u,d,s}\overline{q}_i \hs q_i)\rangle_{\text{inv.}}+\frac{2 x }{9  w^{17/9}}(\langle\overline{q}_d\hs q_d+\overline{q}_u\hs q_d\rangle_{\text{inv.}}\nonumber\\&+\frac{\langle \overline{q}_i q_j+\overline{q}_j q_j\rangle_{\text{inv.}}}{81  w^{17/9}} (w (145 x-81)+128 x \log (w)-523 x) -\frac{ \langle \frac{\as}{\pi}G^2\rangle_{\text{inv.}} }{1296  w^{17/9}}(2 w (145 x-81)+256 x \log (w)-1181 x)\nonumber\\&+\frac{1}{1512 \pi ^2  w^{11/3} x}\Bigg\lbrace \left(m_{j}^4+ m_{j}^4 \right)(-324 w^2 - 189 w x)+m_{i}^2 m_{j}^2 \left(x^2 (-6090 (w-1)-5376 \log (w))+1134 w x\right)\nonumber\\&+\left(m_{i}^3 m_{j}+ m_{i}m_{j}^3\right)\bigg(x \left(-3480 w^2+5073 w-3072 w \log (w)\right)+648 w^2+x^2 (-8555 w-7552 \log (w)+1877))\bigg)\nonumber\\& -81 w x \sum_{k=u,d,s} m_k^4 \Bigg\rbrace\,.
\end{align}
Corresponding FOPT expression is given by:
\begin{align}
   \Psi_4=&\frac{1}{6}((6 L+11) x+3) \langle \sum_{i=u,d,s}\overline{q}_i \hs q_i\rangle_{\text{inv.}}-\frac{1}{3}(2 (3 L+7) x+3)  (\langle \overline{q}_i q_j+\overline{q}_j q_j\rangle_{\text{inv.}})+\frac{1}{9}2 x  \langle\overline{q}_d\hs q_d+\overline{q}_u\hs q_d\rangle_{\text{inv.}}\nonumber\\&+\frac{1}{16}\langle \frac{\as}{\pi}G^2\rangle_{\text{inv.}}((4 L+11) x+2) -\frac{1}{56 \pi ^2 x} \Big(3x \sum_{k=u,d,s} m_k^4-m_i^3 m_j ((90 L+59) x+24)-m_i m_j^3 ((90 L+59) x+24)\nonumber\\&\bs\bs\bs-42 x m_i^2 m_j^2+m_i^4 ((45 L+7) x+12)+m_j^4 (45 L x+7 x+12) \Big)
\end{align}

	\end{widetext}

\end{document}